\documentclass[twocolumn, times]{aastex63}

\usepackage{rotating}
\usepackage{booktabs}
\usepackage[version=4]{mhchem}
\usepackage{multirow}
\usepackage{longtable}
\usepackage{graphicx}
\usepackage{color}
\usepackage{float}
\usepackage{xcolor}
\newcommand{\tr}[1]{\textcolor{black}{#1}}

\begin{document}

\title{\large{A Next-Generation Exoplanet Atmospheric Retrieval Framework \texttt{NEXOTRANS} for Emission Spectroscopy: New Constraints and Atmospheric Characterization of WASP-69b Using JWST NIRCam and MIRI Observations}}

\correspondingauthor{Liton Majumdar}
\email{liton@niser.ac.in, dr.liton.majumdar@gmail.com}

\author[0009-0007-7880-0250]{Tonmoy Deka}
\affiliation{Exoplanets and Planetary Formation Group, School of Earth and Planetary Sciences, National Institute of Science Education and Research, Jatni 752050, Odisha, India}
\affiliation{Homi Bhabha National Institute, Training School Complex, Anushaktinagar, Mumbai 400094, India}

\author[0000-0001-7031-8039]{Liton Majumdar}
\affiliation{Exoplanets and Planetary Formation Group, School of Earth and Planetary Sciences, National Institute of Science Education and Research, Jatni 752050, Odisha, India}
\affiliation{Homi Bhabha National Institute, Training School Complex, Anushaktinagar, Mumbai 400094, India}

\author[0009-0001-3342-3996]{Tasneem Basra Khan}
\affiliation{Exoplanets and Planetary Formation Group, School of Earth and Planetary Sciences, National Institute of Science Education and Research, Jatni 752050, Odisha, India}
\affiliation{Homi Bhabha National Institute, Training School Complex, Anushaktinagar, Mumbai 400094, India}

\author[0009-0002-6571-9406]{Swastik Dewan}
\affiliation{Exoplanets and Planetary Formation Group, School of Earth and Planetary Sciences, National Institute of Science Education and Research, Jatni 752050, Odisha, India}
\affiliation{Homi Bhabha National Institute, Training School Complex, Anushaktinagar, Mumbai 400094, India}

\author[0009-0002-4995-9346]{Priyankush Ghosh}
\affiliation{Exoplanets and Planetary Formation Group, School of Earth and Planetary Sciences, National Institute of Science Education and Research, Jatni 752050, Odisha, India}
\affiliation{Homi Bhabha National Institute, Training School Complex, Anushaktinagar, Mumbai 400094, India}

\author[0009-0001-4100-9218]{Debayan Das }
\affiliation{Exoplanets and Planetary Formation Group, School of Earth and Planetary Sciences, National Institute of Science Education and Research, Jatni 752050, Odisha, India}
\affiliation{Homi Bhabha National Institute, Training School Complex, Anushaktinagar, Mumbai 400094, India}

\author[0009-0002-6571-9406]{Mithun Patra}
\affiliation{Exoplanets and Planetary Formation Group, School of Earth and Planetary Sciences, National Institute of Science Education and Research, Jatni 752050, Odisha, India}
\affiliation{Homi Bhabha National Institute, Training School Complex, Anushaktinagar, Mumbai 400094, India}


\begin{abstract}

Thermal emission spectra provide key insights into the atmospheric composition and especially the temperature structure of an exoplanet. With broader wavelength coverage, sensitivity and higher resolution, JWST has enabled robust constraints on these properties, including detections of photochemical products. This advances the need for retrieval frameworks capable of navigating complex parameter spaces for accurate data interpretation. In this work, we introduce the emission retrieval module of \texttt{NEXOTRANS}, which employs both one- and two-stream radiative transfer approximations and leverages Bayesian and machine learning techniques for retrievals. It also incorporates approximate disequilibrium chemistry models to infer photochemical species like SO$_2$. We applied \texttt{NEXOTRANS} to the JWST NIRCam and MIRI emission observations of WASP-69b, covering the 2–12 $\mu$m range. The retrievals place robust constraints on the volume mixing ratios (VMR) of H$_2$O, CO$_2$, CO, CH$_4$, and potential SO$_2$. The best-fit model, i.e, free chemistry combined with non-uniform aerosol coverage, yields a log(VMR) = $-3.78^{+0.15}_{-0.17}$ for H$_2$O and $-5.77^{+0.09}_{-0.10}$ for CO$_2$ which has a sharp absorption at 4.3 $\mu$m. The second best-fit model, the hybrid equilibrium chemistry (utilizing equilibrium chemistry-grids) combined with non-uniform aerosol yields a C/O of $0.42^{+0.17}_{-0.13}$ and a metallicity of log[M/H] = $1.24^{+0.17}_{-0.14}$, corresponding to approximately 17.38 times the solar value. 
This hybrid chemistry retrieval also constrain SO$_2$ with a log(VMR) = $-4.85^{+0.28}_{-0.29}$, indicating possible absorption features in the 7--8~$\mu$m range. These results highlight \texttt{NEXOTRANS}'s capability to significantly advance JWST emission spectra interpretation, offering broader insights into exoplanetary atmospheres.

\end{abstract}


\section{\textbf{Introduction}} \label{sec:intro}

Observing a planet’s own radiation rather than the filtered starlight as in transmission, emission spectroscopy allows constraints on the temperature-pressure profile, atmospheric circulation, and energy-transport mechanisms \citep{mansfield2023revealing}. In contrast to transmission spectroscopy, which primarily probes the terminator region and is more susceptible to clouds and hazes obscuring deeper layers, emission spectroscopy directly measures outgoing planetary flux, revealing deeper atmospheric layers and temperature-pressure profiles. This distinction makes emission retrievals essential for characterizing temperature structure, thermal dissociation effects, and metallicity variations in exoplanetary atmospheres, particularly for hot and ultra-hot Jupiters where strong thermal emissions dominate. Emission spectra also serve as a crucial diagnostic tool for probing the thermal structure of exoplanetary atmospheres, particularly for identifying thermal inversion-regions where temperature increases with altitude \citep{fortney2018modeling}. In the absence of inversion, molecular bands typically appear in absorption, as radiation from deeper, hotter layers is absorbed by the cooler upper layers. However, when thermal inversions are present, molecules such as CO, CO$_2$, H$_2$O, VO and TiO produce emission features instead, as these species become optically thick at higher, hotter layers \citep{madhusudhan2010inference}. The presence of TiO and VO, in particular, has been suggested as a key indicator of thermal inversions in hot Jupiter atmospheres, as these species can act as strong optical absorbers, deposit stellar irradiation at high altitudes and drive temperature inversions \citep{gandhi2019new}. By analyzing the emission spectra, one can directly infer the temperature gradient of the atmosphere and identify the chemical processes governing the energy balance \citep{Madhusudhan_2019}.

Before JWST, space telescopes such as the Hubble Space Telescope (HST) and the Spitzer Space Telescope played a pivotal role in advancing our understanding of exoplanetary atmospheres through emission spectroscopy \citep{Deming2005, deming2006strong, grillmair2007spitzer, charbonneau2008broadband, stevenson2014deciphering, deming2020highlights}. Hubble’s powerful suite of instruments--including the Cosmic Origins Spectrograph (COS), and the Wide Field Camera 3 (WFC3) enabled detailed atmospheric characterization \citep{France_2010, foote2021emission}. HST’s WFC3 has been extensively used to probe thermal emission in the near-infrared (1.1–1.6~$\mu$m), leading to the detection of water vapor (H$_2$O) in numerous hot Jupiters, such as WASP-121b \citep{mikal2019emission}. Retrieval analyses of these datasets have revealed thermal inversions in some exoplanets, attributed to high-altitude absorbers such as TiO and VO \citep{changeat2021hubble}. The emission spectrum of CoRoT-1b, obtained with Spitzer and HST, indicated inefficient heat redistribution and suggested subsolar metallicity with a low C/O ratio \citep{glidic2022atmospheric}. Similarly, WASP-103b’s thermal emission spectrum hinted at either a thermally inverted atmosphere or an isothermal featureless spectrum, although existing data lack the resolution to definitively distinguish between these scenarios. Additional transit observations in the optical and NIR regions could clarify whether the atmosphere is truly isothermal or whether clouds and haze create a pseudo-isothermal effect, whereas the detection of TiO or CH$_4$ could offer critical insights into its thermal structure and composition \citep{cartier2016near}. The limited wavelength coverage of these instruments has often hindered comprehensive atmospheric constraints, necessitating complementary ground-based spectroscopy, which typically requires several pre-processing steps such as telluric correction \citep{birkby2018exoplanet}.

Building upon these foundations, the JWST’s advanced instrumentation has provided a transformative leap in exoplanetary atmospheric characterization, offering higher spectral resolution, sensitivity and wavelength coverage. New observations of GJ 436b with JWST have provided a robust constraint on its flux, temperature, metallicity, C/O, and evidence for CO$_2$ compared to previous observations from Spitzer and HST \citep{mukherjee2025jwst}. JWST emission observations of WASP-77Ab enabled precise measurements of its subsolar metallicity and enabled a comparison of its overall properties with those of other hot Jupiters, establishing diversity among hot Jupiters \citep{august2023confirmation}. Additionally, with JWST, the era of emission spectroscopy of terrestrial exoplanets has just begun \citep{greene2023thermal}, and the influence of host stars on their atmospheres. The thermal mission spectrum of LTT 1445A b, a rocky exoplanet observed with JWST MIRI/LRS, lacks a thick atmosphere, suggesting erosion around M-dwarf systems \citep{wachiraphan2024thermal}. MIRI observations of the terrestrial exoplanet GJ 1132b \citep{xue2024jwstemission} obtained an emission spectrum which was consistent with a featureless blackbody, suggesting that GJ 1132b likely does not have a significant atmosphere, supporting the concept of a universal `cosmic shoreline' given the high level of bolometric and extreme ultraviolet (EUV) and X-ray irradiation received by the planet. Hence, the JWST has been successful in conducting studies that provide valuable insights into the nature of rocky planets orbiting M dwarf stars and their potential to retain an atmosphere.

Numerous exoplanet atmospheric retrieval algorithms are available for analyzing thermal emission spectra \citep{madhusudhan2009temperature,lee2012optimal, line2013systematic, waldmann2015tau, lavie2017helios, gandhi2018retrieval, molliere2019petitradtrans, min2020arcis, kitzmann2020helios, cubillos2021pyrat, kawahara2022exojax, robinson2023exploring, macdonald2023catalog}, utilizing a range of methodologies, including parametric as well as self-consistent radiative–convective equilibrium models. This work extends the \texttt{NEXOTRANS} retrieval framework introduced in \citet{deka2025} to include emission spectroscopy, thereby facilitating a more comprehensive characterization of exoplanetary atmospheres. To ensure the robustness of the retrievals, both the one-stream radiative transfer approximation—which provides a simplified yet computationally efficient approach—and a more detailed two-stream radiative transfer method for enhanced accuracy were implemented. The flexible, modified hybrid and equilibrium offset chemistry approaches of \texttt{NEXOTRANS} were also employed to infer the presence of disequilibrium processes such as photochemistry. The \texttt{NEXOTRANS} emission retrievals were applied to JWST observations of WASP-69b, obtained with NIRCam and MIRI \citep{schlawin2024multiple}, covering a broad wavelength range of 2.0--12.0~$\mu$m. 
For parameter estimation, the \texttt{PyMultiNest} Bayesian nested sampling framework was utilized alongside a machine learning framework using the stacking regressor algorithm, ensuring a thorough, efficient, and independent exploration of the parameter space. This capability of \texttt{NEXOTRANS} to perform retrievals using diverse approaches and models highlights its potential to contribute meaningfully to the study of exoplanet atmospheres using emission spectra.

In selecting a suitable target for our study, we prioritized exoplanets with spectroscopic observations covering a broad wavelength range, as such coverage is essential for constraining atmospheric composition with high confidence. The availability of WASP-69 b’s emission spectrum from 2–12 $\mu$m provided an excellent opportunity, since shorter wavelength baselines alone are often insufficient to robustly determine key molecular abundances. Moreover, \citet{schlawin2024multiple} reported evidence of photochemically produced SO$_2$ in WASP-69 b’s atmosphere, motivating us to explore this possibility using our approximate disequilibrium chemistry frameworks such as the hybrid equilibrium approach.


\citet{schlawin2024multiple} analyzed JWST observations of the hot Saturn-mass exoplanet WASP-69 b by combining two NIRCam grism time-series datasets spanning 2.4–5.0~$\mu$m with a MIRI low-resolution spectrometer (LRS) dataset covering 5–12~$\mu$m. A homogeneous one-region model provided a poor fit, whereas a two-region model with distinct temperature–pressure (TP) profiles and cloud properties for the hot and cool portions of the dayside yielded a significantly improved match to the observed spectrum. Additional models incorporating either a high, wavelength-independent geometric albedo (scattering model) or a high-altitude silicate cloud deck (cloud-layer model) also reproduced key spectral features. The retrieved TP profiles consistently indicated an inhomogeneous dayside, with a hotter region covering approximately 68\% of the surface and a cooler, cloudier region comprising the remainder, consistent with inefficient day-night heat redistribution. Chemically, the emission spectrum exhibited strong absorption features from H$_2$O, CO, and CO$_2$, but no evidence of CH$_4$, despite its predicted abundance at the equilibrium temperature of 963~K. The retrievals indicated a supersolar metallicity, with the two-region and cloud-layer models favoring enrichments of 6–14$\times$ solar and carbon-to-oxygen ratios (C/O) of 0.65–0.94, while the scattering model yielded somewhat lower values of 4–8$\times$ solar and C/O ratios of 0.26–0.58. Overall, these results suggest a metal-enriched atmosphere containing aerosols, either highly reflective particles or high-altitude silicate clouds, and inefficient redistribution of heat from the dayside to the nightside.


The remainder of this paper is organized as follows: Section \ref{subsec: forward model} introduces the forward modeling including the radiative transfer framework and atmospheric profiles implemented in \texttt{NEXOTRANS}. Section \ref{subsection:retrieval framework} outlines the Bayesian nested sampling and machine learning retrieval techniques employed. In Section \ref{validation_section}, we present the benchmark results of the \texttt{NEXOTRANS} emission module against \texttt{POSEIDON}. Section \ref{results} details the retrieval outcomes and provides constraints on key atmospheric parameters. In Section \ref{discussion}, we examine the best-fit model and discuss the implications of these results on WASP-69b's atmosphere. Finally, Section \ref{sec:conc} conclude the paper with a summary of atmospheric inferences for WASP-69b.

\section{\textbf{THE NEXOTRANS EMISSION RETRIEVAL FRAMEWORK AND ITS APPLICATION}} \label{sec: retrieval framework} 

\begin{figure*}
    \centering
    \includegraphics[width=1.0\linewidth]{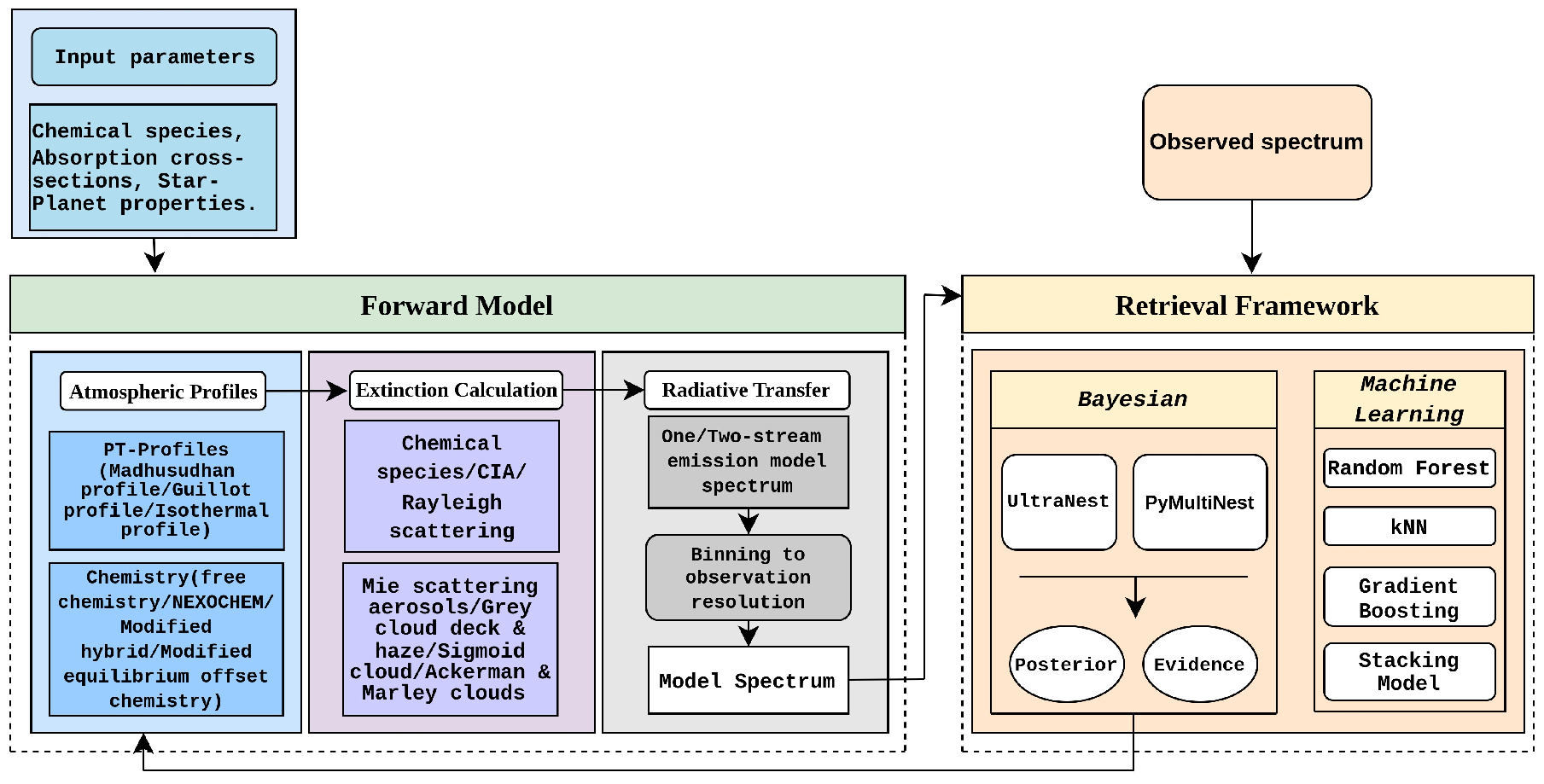}
    \caption{A schematic representation of the retrieval framework implemented in the \texttt{NEXOTRANS} emission module is shown. This framework comprises two key components: the Forward Model, which includes both one-stream and two-stream radiative transfer approximations, and the Retrieval Framework. The Forward Model simulates the exoplanet's atmosphere to generate a model emission spectrum, while Bayesian inference and machine learning techniques are utilized for reliable parameter estimation.}
    \label{fig:flowchart}
\end{figure*}

The emission spectrum retrieval in NEXOTRANS combines a parametric forward model with a retrieval framework, as illustrated in Figure \ref{fig:flowchart}. The emission retrievals in NEXOTRANS follows the same methodology outlined in \citet{deka2025}, employing both nested sampling algorithms (PyMultiNest/UltraNest) and machine learning (Stacking Regressor) for comparative retrieval. To simulate the emission spectra, we introduce a new forward model that incorporates both single-stream and two-stream radiative transfer approximations. The following sections provide a detailed discussion of the radiative transfer framework and key model parameters.

\subsection{\textbf{THE FORWARD MODEL}} \label{subsec: forward model}
To compute the thermal emission spectrum of a planet, the key quantity required is the emergent flux at the top of the atmosphere. This flux is determined by solving the radiative transfer equation, which governs the propagation of radiation through the atmospheric layers. Because an exact solution is often computationally expensive, various approximations are employed to simplify the problem. Two widely used approaches are the one-stream \citep{gandhi2018retrieval} and two-stream approximations \citep{toon1989rapid}.

\texttt{NEXOTRANS} incorporates both these methods to provide a more comprehensive and comparative analysis of the emission spectra. In the following sections, we briefly outline the principles of these two approximations.

\subsubsection{\textbf{One-stream Approximation}} \label{subsubsec: radiative transfer}

In the one-stream approximation \citep{gandhi2018retrieval}, we consider the radiative transfer solution in the pure absorption limit, where no scattering due to the atmosphere is assumed. The intensity emerging out of an atmospheric layer with optical depth $\tau$ and temperature, T, is given by

\begin{equation}
    I_1(\lambda,\mu) = I_0(\lambda,\mu)e^{-\tau/\mu} + B(T,\lambda)(1 - e^{-\tau/\mu})
\end{equation}

where, $I_0$ is the intensity emerging from the lower atmospheric layer at an angle $\theta$ to the normal, with $\mu$ = cos($\theta$). $B(T,\lambda)$ is the Planck function at a temperature $T$ and wavelength, $\lambda$. The intensity at the lowest atmospheric layer is assumed to follow blackbody radiation, with B(T[0], $\lambda$), where T[0] is the temperature of the lowest layer.

The emergent intensity at the top of the atmosphere is obtained by integrating the contributions from all underlying layers along the line of sight. \texttt{NEXOTRANS} uses a Gaussian quadrature with three discrete angles ($\mu$ = $\frac{1}{2}-\frac{1}{2}\sqrt{\frac{3}{5}}, \frac{1}{2}, \frac{1}{2}+\frac{1}{2}\sqrt{\frac{3}{5}}$), each weighted by the corresponding Gaussian weights  5/18, 4/9, 5/18 to efficiently capture angular dependence and ensure accurate flux calculations. Therefore, the flux exiting at the top of atmosphere is given by

\begin{equation}
    F_{top}(\lambda) = 2\pi\int_0^1\mu I_{top}(\lambda,\mu)d\mu
\end{equation}

where, $I_{top}$ is the outgoing intensity at the top of the atmosphere.

Additionally, the planet-star flux ratio can be calculated as

\begin{equation}
    \text{Flux ratio} = \left(\frac{F_{top}}{F_*}\right)\left(\frac{R_p}{R_*}\right)^2
\end{equation}

Here, \( F_* \) represents the stellar flux, which can be approximated as a blackbody spectrum given by \( B_*(T_*, \lambda) \), where \( T_* \) is the effective temperature of the star. R$_p$ and R$_*$ are the planetary and stellar radii respectively. Alternatively, \( B_* \) can be replaced with more detailed stellar spectrum models, such as the PHOENIX \citep{husser2013new} and Kurucz models \citep{kurucz1975table, castelli2004new}.

The simplified one/single-stream approximation described above becomes inaccurate once the effects of strong scattering due to clouds or aerosols comes into place \citep{toon1989rapid}.
In this case, the contribution of scattering to the total extinction increases, thereby altering the transport of radiation through the atmosphere. The one-stream method, which only considers radiation propagating in one direction with pure absorption, fails to capture the effects of multiple scattering, leading to inaccuracies in both flux calculations and inferred atmospheric properties \citep{de2011influence}. By explicitly accounting for both upward and downward flux components, the two-stream approximation provides a more accurate solution, particularly in atmospheres in which clouds, aerosols, or Rayleigh scattering plays a significant role.

Therefore, \texttt{NEXOTRANS} also incorporates the two-stream radiative transfer method to model the emission spectrum. We discuss this in the next section.

\subsubsection{\textbf{Two-stream Approximation}}

In the most general case, solving the radiative transfer equation requires tracking radiation at all possible angles across both the zenith and azimuthal directions, making computations extremely expensive when integrating over many discrete directions. Even if the azimuthal dependence is neglected, the remaining angular dependencies still pose significant challenges, particularly when scattering is involved. The two-stream approximation simplifies this by considering only two directions, an upward and a downward stream, eliminating the need to resolve the full angular distribution of intensity. In \texttt{NEXOTRANS}, we adopt the two-stream radiative transfer methodology from \citet{toon1989rapid} to compute the radiative fluxes.

The general equation of radiative transfer in a plane parallel scattering atmosphere is given by

\begin{equation}\label{RTeq}
\begin{gathered}
\mu \frac{\partial I_v}{\partial \tau_v}\left(\tau_v, \mu, \phi\right)=I_v\left(\tau_v, \mu, \phi\right)-S_v\left(\tau_v, \mu, \phi\right)-
\\
\frac{\omega_{0 v}}{4 \pi} 
\int_0^{2 \pi} \int_{-1}^1 P_v\left(\mu, \mu^{\prime}, \phi, \phi^{\prime}\right) I_v\left(\tau_v, \mu^{\prime}, \phi^{\prime}\right) d \mu^{\prime} d \phi^{\prime}
\end{gathered}
\end{equation}

where \( \mu \) is the cosine of the angle at which the intensity \( I_{\nu} \) is observed, measured relative to the surface normal; \( \tau \) is the optical depth; \( \omega_0 \) is the single scattering albedo, representing the fraction of extinction due to scattering; \( P \) is the scattering phase function, describing the angular distribution of scattered radiation; \( \nu \) is the frequency; and \( S_{\nu} \) is the source function, which accounts for the atmospheric emission and scattered radiation.

Following \citet{toon1989rapid}, the azimuthally integrated upward (downward) flux is,

\begin{equation}
    F^{\pm}_{\nu} = \int_0^{1}\mu I_\nu^\pm(\tau, \mu)d\mu
\end{equation}

where, the specific intensities $I_\nu(\tau_\nu,\mu, \phi)$ are integrated azimuthally to calculate $I_\nu(\mu,\tau_\nu)$.

Integrating Equation \ref{RTeq}, any two-stream expression can be written in terms of two coupled equations as follows:-

\begin{equation}
\begin{gathered}
    \frac{\partial F^+_\nu}{\partial \tau_\nu} = \gamma_1F^+_\nu - \gamma_2F^-_\nu - S^+_\nu
    \\
    \frac{\partial F^+_\nu}{\partial \tau_\nu} = \gamma_2F^+_\nu - \gamma_1F^-_\nu + S^-_\nu
    \end{gathered}
\end{equation}

where \( \gamma_1 \) and \( \gamma_2 \) are functions of the scattering properties of the medium and depend on the specific form of the two-stream approximation. In our case, we adopt the Hemispheric Mean approximation, where  

\begin{equation}
\gamma_1 = 2 - \omega_0 (1 + g), \quad \gamma_2 = \omega_0 (1 - g),
\end{equation}

where \( \omega_0 \) and \( g \) represent the single scattering albedo and scattering asymmetry parameters of the atmospheric layer, respectively.

The source functions for the upward and downward intensities are written as follows,

\begin{equation}
    S^+_\nu = Ge^{\lambda t} + He^{-\lambda t} + \alpha_1 + \alpha_2\tau
\end{equation}

\begin{equation}
    S^-_\nu = Je^{\lambda t} + Ke^{-\lambda t} + \sigma_1 + \sigma_2\tau
\end{equation}

where the quantities G, H, J, K, \(\alpha_1, \alpha_2, \sigma_1 \) and \(\sigma_2\) are the parameters in hemispheric mean two-stream source function technique presented in \citet{toon1989rapid}.

With the source function specified, the upward azimuthally averaged intensity at the top of the layer is given as

\begin{equation}\label{twostream}
\begin{gathered}
        I_n^+(\tau=0,\mu) = I_n^+(\tau,\mu)e^{-\tau/\mu}
        \\
        + \frac{G}{(\lambda\mu-1)}[e^{-\tau/\mu} - e^{-\tau\lambda}]
        \\
        + \frac{H}{(\lambda\mu+1)}[1-e^{-\tau(\lambda+1/\mu)}]
        \\
        + \alpha_1[1-e^{-\tau/\mu}] + \alpha_2[\mu-(\tau+\mu)e^{-\tau/\mu}]
\end{gathered}
\end{equation}

The final outgoing thermal flux is then obtained by calculating this intensity at five different emergent angles, with angle cosines given by $\mu$ = 0.0985, 0.3045, 0.5620, 0.8019, 0.9601 and weights = 0.0157, 0.0739, 0.1463, 0.1671, 0.0967 \citep{mukherjee2023picaso}.

We impose the following boundary conditions at the top and bottom of the atmosphere to initiate the calculation  of Equation \ref{twostream},

\begin{equation}
\begin{gathered}
    B_{top} = 0
    \\
    B_{bot} = B(T_{{bot}}) + \mu_1 \frac{B(T_{bot-1} ) - B(T_{bot})}{\tau} 
\end{gathered}
\end{equation}

where $B(T_{bot}$) represents the blackbody function at temperature, $T_{bot}$, at the lowest layer of the atmosphere, $\tau$ is the optical depth of the lowest layer and $\mu_1$ is 0.5 following the hemispheric-mean approximation.

\begin{figure*}[]
    \centering
    \includegraphics[width=1\linewidth]{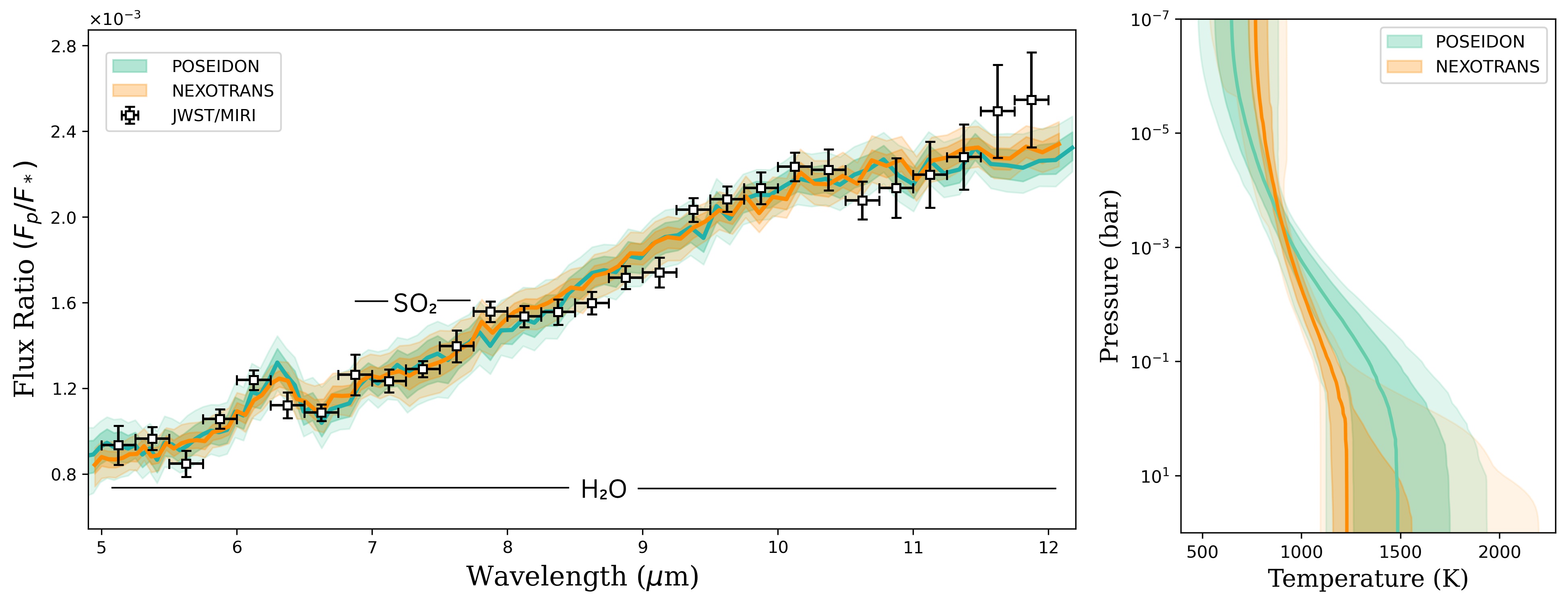}
    \parbox{0.8\linewidth}{\centering (a) \texttt{NEXOTRANS} one-stream validation with \texttt{POSEIDON}.}
    \label{fig:spectrum}
    \includegraphics[width=1\linewidth]{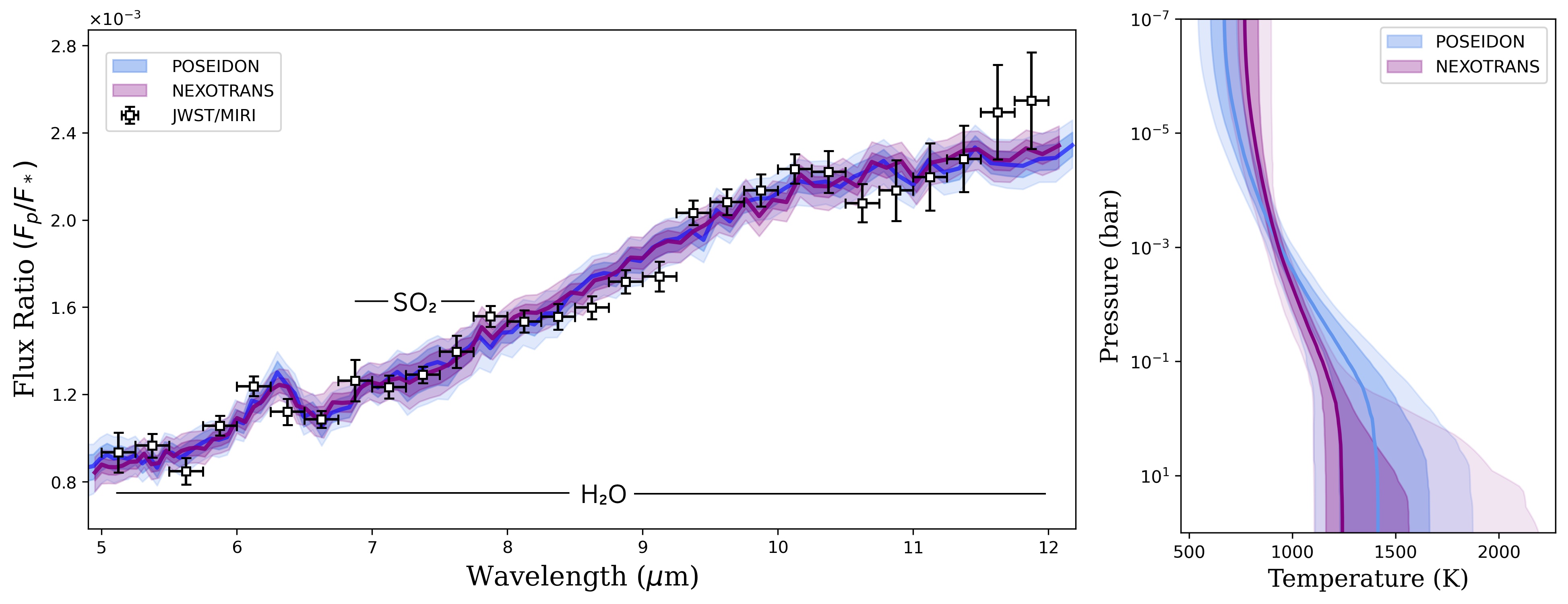}
    \parbox{0.8\linewidth}{\centering (b) \texttt{NEXOTRANS} two-stream validation with \texttt{POSEIDON}.}
    \label{fig:hybrid_ML_spectra}
    \caption{Validation of NEXOTRANS’s emission retrieval module against POSEIDON: (a) Best-fit retrieved spectrum and P-T profile using the one-stream approximation, and (b) using the two-stream approximation. JWST MIRI observations are shown with black error bars. Retrievals were performed with a model resolution of 15,000 and 1,000 live points in the nested sampler. For clarity, the best-fit spectra are shown binned to a resolution of 100, with the median, 1$\sigma$ and 2$\sigma$ confidence intervals. The one-stream approximation assumes a clear atmosphere, while the two-stream approximation includes uniform MgSiO$_3$ aerosol. Both retrievals are performed assuming free chemistry.}
    \label{fig:benchmark_spectra}
\end{figure*}

\begin{table*}
\centering
\caption{Retrieved parameters from the validation of the NEXOTRANS emission retrieval module, compared with POSEIDON, using the WASP-69b JWST MIRI (5-12 $\mu$m) dataset.}
\resizebox{0.70\textwidth}{!}{
\hspace{-2.0cm}
\begin{tabular}{lcccccc}
\toprule
\hline
\multirow{2}{*}{}  & \multicolumn{1}{c}{log(H$_2$O)} & \multicolumn{1}{c}{log(CO$_2$)} & \multicolumn{1}{c}{log(CO)} & \multicolumn{1}{c}{log(SO$_2$)} & \multicolumn{1}{c}{log(CH$_4$)} & \multicolumn{1}{c}{Red.$\chi^2$}\\
\bottomrule 
 & \multicolumn{6}{c}{{\textbf{One-stream approximation (clear atmosphere)}}} \\
\hline
NEXOTRANS &  $-1.73^{+0.46}_{-0.64}$ & $-8.42^{+2.17}_{-2.15}$ & $-7.08^{+3.11}_{-3.05}$ & $-6.41^{+0.67}_{-2.67}$ & $-8.34^{+2.40}_{-2.27}$ & 2.32\\
[0.2cm]
POSEIDON &  $-1.71^{+0.46}_{-0.64}$ & $-8.30^{+2.25}_{-2.20}$ & $-6.95^{+3.17}_{-3.07}$ & $-7.30^{+1.36}_{-2.99}$ & $-8.84^{+2.01}_{-1.97}$ &  2.36\\
[0.2cm]
\hline
 & \multicolumn{6}{c}{{\textbf{Two-Stream approximation (uniform aerosol atmosphere)}}} \\
\hline
NEXOTRANS  &  $-1.64^{+0.40}_{-0.60}$ & $-8.22^{+2.56}_{-2.16}$ & $-6.90^{+3.12}_{-3.02}$ & $-6.33^{+0.64}_{-2.52}$ & $-8.30^{+2.44}_{-2.25}$ & 3.14\\
[0.2cm]
POSEIDON  &  $-1.83^{+0.45}_{-0.40}$ & $-8.23^{+1.96}_{-2.15}$ & $-6.97^{+2.85}_{-2.91}$ & $-6.78^{+0.98}_{-2.95}$ & $-8.51^{+1.91}_{-1.91}$ & 3.24\\
[0.2cm]
\hline
\end{tabular}
}
\label{validation_table}
\end{table*}

\subsubsection{\textbf{Atmospheric Profiles}} \label{subsubsection : atmospheric profiles} 


The accurate modeling of an exoplanetary atmosphere requires both a well-defined pressure-temperature (P-T) profile and a corresponding Volume Mixing Ratio (VMR) profile. This section outlines the parameterizations implemented in \texttt{NEXOTRANS} for these atmospheric profiles.  

For the P-T profile, \texttt{NEXOTRANS} implements three parameterizations: an isothermal profile, the \citet{guillot2010radiative} model, and the \citet{madhusudhan2009temperature} profile, each capturing different thermal characteristics of planetary atmospheres. 

The atmospheric chemistry module in \texttt{NEXOTRANS} uses four approaches. The free chemistry model treats the mixing ratios of chemical species as free parameters, assuming constant values with altitude. The equilibrium chemistry model, utilizing the benchmarked \texttt{NEXOCHEM} module \citep{deka2025}, determines molecular abundances as a function of temperature, pressure, C/O ratio, and metallicity, enabling constraints on global atmospheric composition. To speed up the retrievals, we utilized a precomputed grid spanning temperatures from 300 to 4000 K, pressures from $10^{-7}$ to $10^{2}$ bar, C/O ratios between 0.2 and 2.0, and metallicities ([M/H]) ranging from 0.1 to 1000 times solar, calculated using \texttt{NEXOCHEM}. To account for disequilibrium processes, the modified hybrid equilibrium approach \citep{deka2025} combines equilibrium and free chemistry by treating selected species as vertically constant, whereas the modified equilibrium offset \citep{deka2025} method applies a multiplicative adjustment ($\delta)$ to equilibrium abundances to approximate disequilibrium effects. 

Following the determination of the P-T profile and atmospheric chemistry, \texttt{NEXOTRANS} computes the optical depth $\tau$
for each atmospheric layer.  

\begin{figure} []
	\centering
	\hspace{-0.3cm}
	\begin{minipage}{0.45\textwidth}
		\centering
		\includegraphics[width=\textwidth]{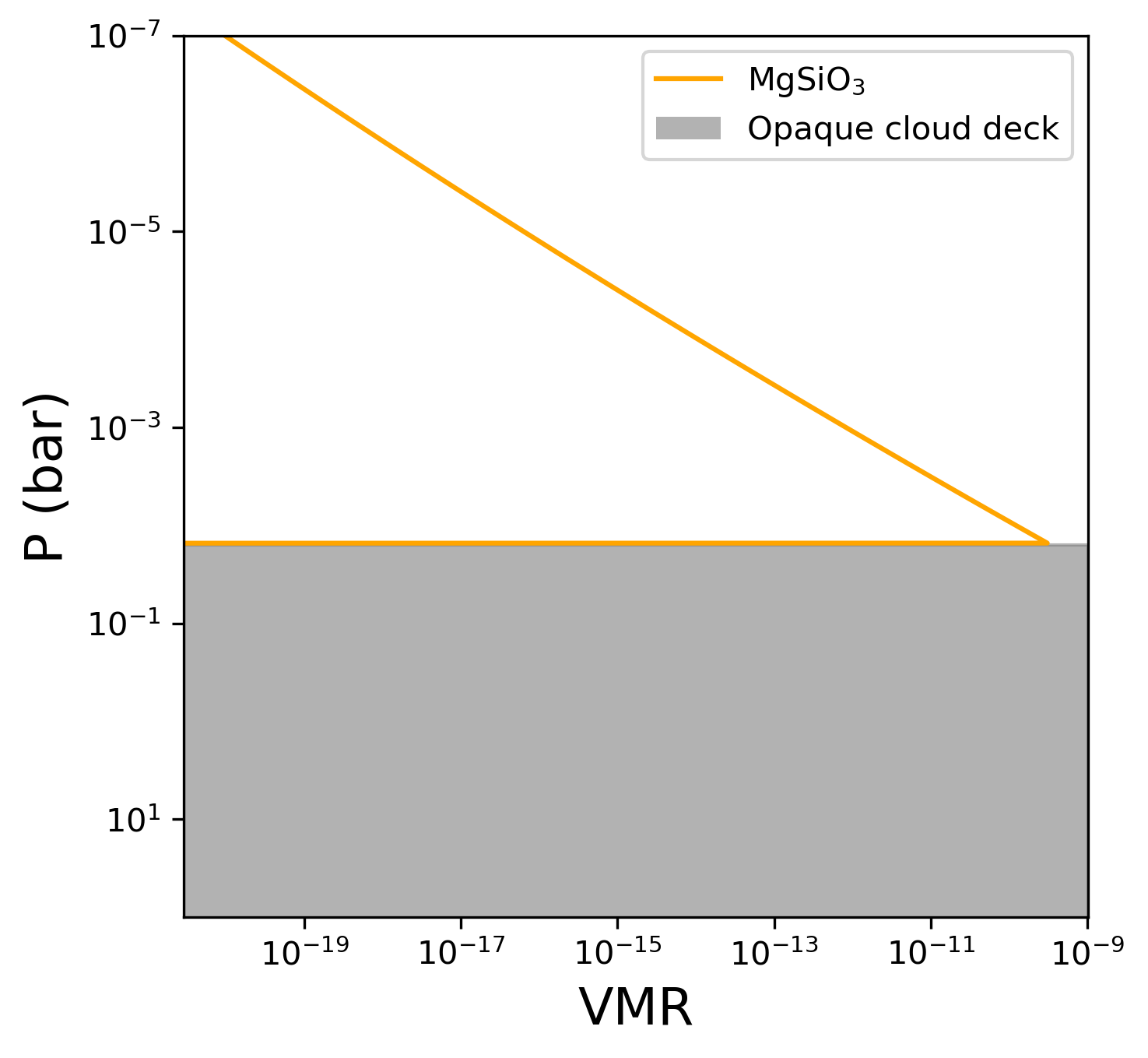}
	\end{minipage}        
	\caption{Vertical mixing ratio profile model of MgSiO$_3$ aerosol as a function of pressure. The shaded gray region represents the opaque cloud deck, extending from the bottom to log(P$_{\mathrm{MgSiO_3,deck}}$). The slanted yellow line denotes the volume mixing ratio (VMR) profile of the mie scattering aerosol.}
	\label{fig:cloud_vmr}
\end{figure}

\subsubsection{\textbf{Opacity Sources}} \label{subsubsection : atmospheric extinction}


The primary sources that shape atmospheric spectra include molecular and atomic absorption, collision-induced absorption (CIA), and scattering processes. The chemical opacity sources incorporated in \texttt{NEXOTRANS} are derived from the open-source absorption cross-section of POSEIDON opacity database\footnote{\url{https://poseidon-retrievals.readthedocs.io/en/latest/content/opacity_database.html}}, ensuring comprehensive coverage of key molecular species.

In addition to chemical signatures, the presence of clouds also plays a crucial role in shaping the atmospheric spectra. \texttt{NEXOTRANS} incorporates both gray and non-gray cloud treatments to provide a more comprehensive interpretation of the observed atmospheric spectra \citep{deka2025}. In addition to modeling patchy gray clouds and hazes using the \citet{line2016influence} prescription, \texttt{NEXOTRANS} also accounts for more complex non-gray cloud structures, including sigmoid clouds \citep{Constantinou}, Mie-scattering aerosols \citep{pinhas2017signatures}, and the Ackerman-Marley cloud model \citep{ackerman2001precipitating}. Figure~\ref{fig:cloud_vmr} illustrates the aerosol parameterization adopted in this study. The aerosol volume mixing ratio peaks at the top of the opaque cloud deck and decreases with altitude, following an exponential decline determined by a specified scale-height factor.

These diverse extinction sources, spanning molecular absorption, collision-induced absorption, Rayleigh scattering and cloud opacity, collectively influence the observed atmospheric spectra of exoplanets. By incorporating a wide range of chemistry and cloud models, \texttt{NEXOTRANS} enables robust and flexible retrievals, ensuring accurate characterization of exoplanetary atmospheres under varying conditions.


\begin{figure*}[]
    \centering
    \includegraphics[width=0.80\linewidth]{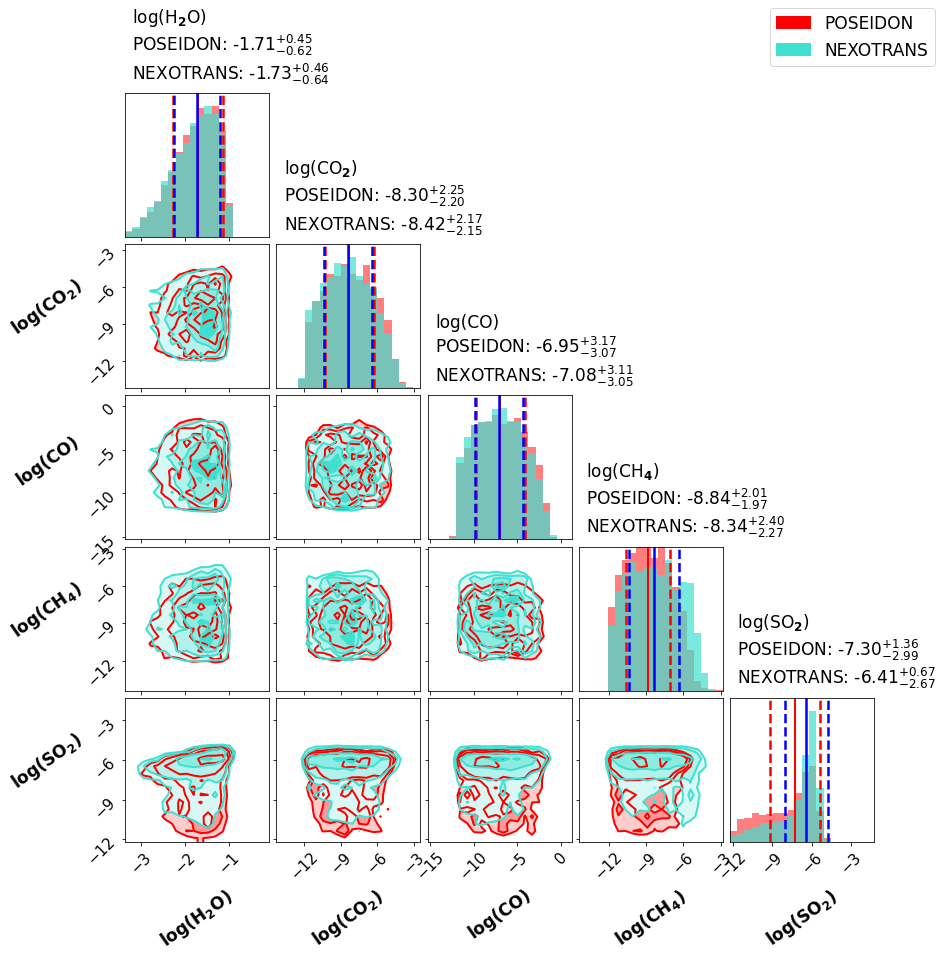}
    \caption{Posterior distributions of free parameters retrieved with the \textit{one-stream radiative transfer} implementation of \texttt{NEXOTRANS}, compared to those obtained with \texttt{POSEIDON}. The retrievals assume a clear atmosphere and free chemistry. The retrieved parameters from both algorithms show agreement within 1$\sigma$, indicating consistency between the two implementations.}
    \label{fig:overplot1}
\end{figure*}

\begin{figure*}[]
    \centering    
    \includegraphics[width=1.0\linewidth]{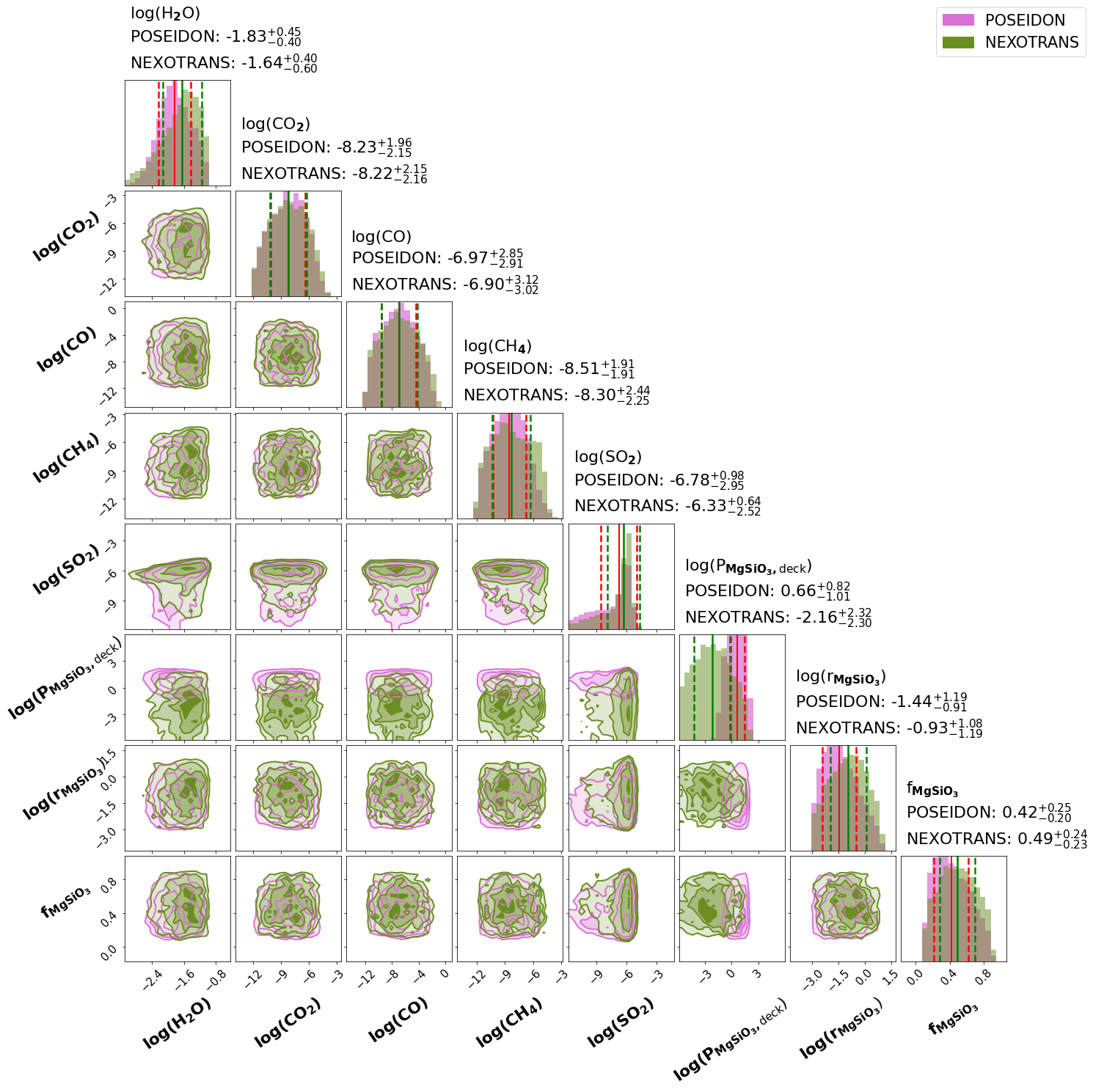}
    \label{fig:hybrid_ML_spectra}
    \caption{Posterior distributions of free parameters retrieved with the \textit{two-stream radiative transfer} implementation of \texttt{NEXOTRANS}, compared to those obtained with \texttt{POSEIDON}. The retrievals assume an atmosphere uniformly covered by MgSiO$_3$ aerosols and adopt a free chemistry framework. The retrieved parameters from both algorithms show agreement within 1$\sigma$, indicating consistency between the two implementations.}
    \label{fig:overplot2}
\end{figure*}


\begin{table*}[t]
\centering
\caption{Free parameters in the retrieval models. The retrieved values are presented in Table \ref{Table:all_params}}
\label{tab:retrieval_models}
\renewcommand{\arraystretch}{1.2} 
\resizebox{1.0\textwidth}{!}{
\begin{tabular}{lc}
\hline
\hline
\textbf{Model}                        & \textbf{Free Parameters}  \\
\hline
\hline
Common Parameters* & log(P$_{\text{ref}}$), $\alpha_1$, $\alpha_2$, log(P$_1$), log(P$_2$), log(P$_3$),  T$_0$, log(MgSiO$_3$),  log(r$_{\text{MgSiO}_3}$), log(P$_{\text{MgSiO}_3,\text{deck}}$), f$_{\text{MgSiO}_3}$, $\phi_{\text{MgSiO}_3}$** \\
\hline
Free Chemistry     & log(H$_2$O), log(CO$_2$), log(CO), log(SO$_2$),  
                                    
                                  log(CH$_4$)  \\
\hline
Equilibrium Chemistry         &  C/O, Metallicity   \\
\hline
Hybrid Equilibrium Chemistry  & C/O, Metallicity, log(SO$_2$)  \\
\hline
Equilibrium Offset Chemistry  & C/O, Metallicity, $\delta$(H$_2$O),  
                                      $\delta$(CO$_2$), $\delta$(CO),  
                                  
                                  $\delta$(CH$_4$), log(SO$_2$)  \\
\hline
\hline
\textbf{Number of Datapoints} & NIRCam: 51,  MIRI: 28 \\
\hline
\hline
\end{tabular}
}
\label{Table:free_parameter}

Notes: The one-stream approximation assumes a clear atmosphere without aerosol scattering; therefore, the MgSiO$_3$ parameters are not included in that case.\\
* Common parameters are included in all retrieval models. \\
** $\phi_{\mathrm{MgSiO}_3}$ is the cloud patchiness parameter and is only present in non-uniform aerosol models. It is fixed to 1 for uniform cloud assumptions. \\

\end{table*}

\subsection{\textbf{RETRIEVAL FRAMEWORK}} \label{subsection:retrieval framework}

The \texttt{NEXOTRANS} emission module follows the same retrieval framework as implemented in the \texttt{NEXOTRANS} transmission module \citep{deka2025}. \texttt{NEXOTRANS} utilizes the Bayesian inference method as well as a machine learning 
algorithm and provides a comprehensive  comparative analysis of both approaches. For Bayesian inference \texttt{NEXOTRANS} employs the \texttt{UltraNest} \citep{buchner2021ultranest} or the \texttt{PyMultiNest} \citep{buchner2014x} nested sampling algorithms that efficiently explores the parameter space and computes posterior distributions as well as the evidence of a model. We adopt \texttt{PyMultiNest} for the retrievals presented in this study, as the extensive model exploration requires an efficient and computationally fast sampler. \texttt{UltraNest}, by contrast, has been found to converge more slowly in similar applications \citep{gebhard2025flow, cp8c-2nbk}.

On the other hand, the machine learning algorithm in \texttt{NEXOTRANS} employs an ensemble learning approach, specifically the \texttt{Stacking Regressor} \citep{deka2025}, which combines \texttt{Random Forest} \citep{breiman2001random}, \texttt{K-Nearest Neighbors} \citep{cover1967nearest}, and \texttt{Gradient Boosting} \citep{friedman2001greedy} as base models, with a ridge regressor serving as the meta-model. By aggregating the strengths of different base models, the ensemble learning method improves retrieval performance while maintaining computational efficiency. All machine learning retrievals in this work are performed with the \texttt{Stacking Regressor} as discussed in \citet{deka2025}.


Since machine learning retrievals generally provide only point predictions, we generate posterior distributions by adding noise to the observed transit depths and then predicting the model outputs using these noisy inputs. This approach yields a distribution of parameters, and the correlations among them can be visualized using corner plots, as discussed in Appendix~\ref{appendix}. The detailed methodology is described in \citet{deka2025}. Although this method of adding noise to the transit depths does not capture the true parameter uncertainties as in Bayesian inference, it still demonstrates how parameters vary with observational noise. While several machine-learning-based retrieval codes exist in the community, as listed in the catalog of \citet{MacDonald_2023}, the machine learning model in \texttt{NEXOTRANS} is specifically designed to combine multiple supervised algorithms into a single prediction framework, allowing comparative analysis using individual algorithms as well. One key challenge that remains is obtaining full posterior distributions without introducing user-defined noise, as discussed earlier and in \citet{deka2025}. Another challenge lies in achieving flexible training, that is, ensuring the model can generalize to variations in the parameter space without requiring complete retraining. Recently, a few machine learning approaches have begun exploring probabilistic frameworks \citep{Gebhard2024, Yip2022, Vasist2023}, which offer a more rigorous treatment of uncertainties by directly learning posterior distributions rather than point estimates. Such an approach will be incorporated into \texttt{NEXOTRANS} in the future.

\subsection{\textbf{VALIDATION OF \texttt{NEXOTRANS} EMISSION MODULE}}
\label{validation_section}

We benchmarked both the one-stream and two-stream emission retrieval modules of \texttt{NEXOTRANS} against \texttt{POSEIDON}\footnote{\url{https://poseidon-retrievals.readthedocs.io/en/latest/index.html}} by performing two sets of retrievals on the JWST MIRI dataset for WASP-69b, covering wavelengths from 5 to 12~$\mu$m. The retrievals employed the \texttt{PyMultiNest} nested sampling method \citep{buchner2014x}, which is the default and computationally less expensive sampler within the \texttt{NEXOTRANS} framework \citep{deka2025}. This is also the same sampler used in \texttt{POSEIDON}. The retrievals were conducted with a model resolution of 15,000 and utilized 1000 live points for the nested sampler. We assumed the presence of chemical species H$_2$O, SO$_2$, CO$_2$, CO, and CH$_4$ under a free chemistry framework with a clear atmosphere for the simplistic one-stream case, whereas the two-stream case included free chemistry with MgSiO$_3$ as a Mie-scattering aerosol.

As illustrated in Figure \ref{fig:benchmark_spectra} (a) and (b), the \texttt{NEXOTRANS} retrievals show good agreement with the \texttt{POSEIDON} one-stream and two-stream retrievals. Table \ref{validation_table} presents the retrieved parameter values for both approximations, demonstrating consistency within 1$\sigma$ across the two methods. Both frameworks yield satisfactory reduced $\chi^2$ values, indicating good fits to the MIRI data. The overplotted corner plots are presented in Figure \ref{fig:overplot1} and \ref{fig:overplot2}. The retrievals identify H$_2$O as the dominant absorber in the 5-12 $\mu$m wavelength range having median log(VMR) values between -1.64 and -1.83, with additional contributions from other species at lower abundances. Moreover, the retrievals suggest potential absorption features due to to trace amounts of the photochemical product SO$_2$, particularly in the 7--8\,$\mu$m region.

For the two-stream approximation, the retrieved aerosol parameters are broadly consistent with one another within 1$\sigma$, as evident from the over-plotted corner plot Figure \ref{fig:overplot2}. \texttt{NEXOTRANS} retrieves a log particle radius of $-0.93^{+1.08}_{-1.19}$ $\mu$m for MgSiO$_3$ present at log pressure levels of $-2.16^{+2.32}_{-2.30}$ bar as compared to \texttt{POSEIDON} values of $-1.44^{+1.19}_{-0.91}$ and $0.65^{+0.82}_{-1.01}$ respectively. 
\texttt{NEXOTRANS} also retrieves a log mixing ratio of $-10.07^{+5.45}_{-5.68}$ for MgSiO$_3$ at the cloud deck level, compared to $-16.20^{+3.80}_{-1.39}$ as retrieved by \texttt{POSEIDON}, withing the lower 1$\sigma$. The retrieved fractional scale height of the aerosol also agrees with each other, with \texttt{NEXOTRANS} and \texttt{POSEIDON} having values $0.49^{+0.24}_{-0.23}$ and $0.42^{+0.25}_{-0.20}$ respectively 


\begin{table*}
\centering
\caption{Retrieved free parameters for chemical species across all explored atmospheric models. The clear and cloudy atmospheres utilize the one-stream and two-stream radiative transfer approximations, respectively. We also report the reduced $\chi^2$ values of Bayesian retrievals along with the residual and R$^2$ Score for ML retrievals corresponding to each model.}
\hspace{-2.0cm}
\resizebox{1.0\textwidth}{!}{
\vspace{1.9cm}
\begin{tabular}{lccccccccccccc}
\toprule
\hline

& \multicolumn{7}{c}{{\textbf{Free Chemistry}}} \\
\hline

\multirow{2}{*}{} & \multicolumn{1}{c}{log(H$_2$O)} & \multicolumn{1}{c}{log(CO$_2$)} & \multicolumn{1}{c}{log(CO)} & \multicolumn{1}{c}{log(CH$_4$)} & \multicolumn{1}{c}{log(SO$_2$)} & & & Red. $\chi^2$ & \multicolumn{1}{c}{ln(Z)} & Avg residual x $10^{-4}$ & R$^2$ Score\\

\bottomrule 

& \multicolumn{6}{c}{\textit{{One-Stream Approximation (Clear)}}} \\
Bayesian & $-4.69^{+0.25}_{-0.22}$ & $-6.03^{+0.37}_{-0.28}$ & $-4.38^{+0.37}_{-0.25}$ & $-10.28^{+1.13}_{-1.09}$ & $-6.93^{+0.45}_{-0.39}$ & & & 3.21 & 596.09\\
[0.2cm]
ML & \tr{$-5.29^{+0.51}_{-0.20}$} & \tr{$-5.99^{+0.41}_{-0.00}$} & \tr{$-4.52^{+0.21}_{-0.24}$} & \tr{$-10.26^{+0.03}_{-0.03}$} & \tr{$-6.13^{+0.08}_{-0.36}$} & & & \tr{3.05} & & \tr{0.859} & 0.992 \\
[0.2cm]
\hline
& \multicolumn{6}{c}{\textit{{Two-Stream Approximation (Uniform Cloud)}}} \\
Bayesian & $-4.42^{+0.22}_{-0.21}$ & $-5.55^{+0.18}_{-0.16}$ & $-3.95^{+0.27}_{-0.25}$ & $-8.17^{+0.30}_{-0.32}$ & $-6.38^{+0.15}_{-0.15}$ & & & 3.51 & 596.86\\
[0.2cm]
ML & \tr{$-4.50^{+0.50}_{-0.00}$} & \tr{$-6.00^{+1.00}_{-0.00}$} & \tr{$-3.28^{+0.05}_{-0.54}$} & \tr{$-8.20^{+0.70}_{-0.00}$} & \tr{$-6.50^{+0.50}_{-0.00}$} & & & \tr{2.29} & & \tr{0.386} & 0.994\\
[0.2cm]
\hline
& \multicolumn{6}{c}{\textit{{Two-Stream Approximation (Non-Uniform Cloud)}}} \\
Bayesian & $-3.78^{+0.15}_{-0.17}$ & $-5.77^{+0.09}_{-0.10}$ & $-3.78^{+0.21}_{-0.22}$ & $-7.85^{+0.16}_{-0.14}$ & $-11.41^{+0.41}_{-0.37}$ & & & 2.37 & 622.77\\
[0.2cm]
ML & \tr{$-4.00^{+0.41}_{-0.00}$} & \tr{$-5.72^{+0.47}_{-0.20}$} & \tr{$-3.18^{+0.16}_{-0.46}$} & \tr{$-7.01^{+0.02}_{-1.00}$} & \tr{$-9.70^{+0.36}_{-0.30}$} & & & \tr{1.40} & & \tr{0.240} & 0.989 \\
[0.2cm]
\hline
\hline

& \multicolumn{7}{c}{{\textbf{Equilibrium Chemistry}}} \\
\hline

\multirow{2}{*}{} & \multicolumn{1}{c}{C/O} & \multicolumn{1}{c}{log[M/H]} & & & & & & \\
\bottomrule

& \multicolumn{6}{c}{\textit{{One-Stream Approximation (Clear)}}} \\ 
Bayesian & $0.29^{+0.07}_{-0.05}$ & $0.92^{+0.06}_{-0.05}$ & & & & & & 4.36 & 548.06\\
[0.2cm]
ML & \tr{$0.23^{+0.06}_{-0.02}$} & \tr{$0.93^{+0.02}_{-0.02}$} & & && & & \tr{2.87} && \tr{0.295} & 0.987\\
[0.2cm]
\hline
& \multicolumn{6}{c}{\textit{{Two-Stream Approximation (Uniform Cloud)}}} \\
Bayesian & $0.30^{+0.06}_{-0.05}$ & $0.93^{+0.06}_{-0.05}$ & & & & & & 4.63 & 551.09\\
[0.2cm]
ML & \tr{$0.37^{+0.03}_{-0.03}$} & \tr{$0.97^{+0.02}_{-0.06}$} &&&& & & \tr{3.56} && \tr{2.90} & 0.840 \\
[0.2cm]
\hline
& \multicolumn{6}{c}{\textit{{Two-Stream Approximation (Non-Uniform Cloud)}}} \\ 
Bayesian & $0.51^{+0.15}_{-0.15}$ & $1.26^{+0.07}_{-0.07}$ & & & & & & 2.75 & 615.50\\ 
[0.2cm]
ML & \tr{$0.52^{+0.01}_{-0.01}$} & \tr{$1.21^{+0.01}_{-0.00}$} && & && & \tr{2.71} & & \tr{0.306} & 0.837 \\
[0.2cm]
\hline
\hline

& \multicolumn{7}{c}{{\textbf{Hybrid Equilibrium}}} \\
\hline

\multirow{2}{*}{} & \multicolumn{1}{c}{C/O} & \multicolumn{1}{c}{log[M/H]} & \multicolumn{1}{c}{log(SO$_2$)} & & && & \\
\bottomrule

& \multicolumn{6}{c}{\textit{{One-Stream Approximation (Clear)}}} \\
Bayesian & $0.56^{+0.08}_{-0.09}$ & $1.27^{+0.12}_{-0.12}$ &  $-5.09^{+0.17}_{-0.19}$&  & & & & 3.76 & 571.78\\ 
[0.2cm]
ML & \tr{$0.57^{+0.03}_{-0.06}$} & \tr{$1.00^{+0.05}_{-0.00}$} & \tr{$-5.00^{+0.00}_{-0.32}$} & & && & \tr{2.57} & & \tr{0.958} & 0.989 \\
[0.2cm]
\hline
& \multicolumn{6}{c}{\textit{{Two-Stream Approximation (Uniform Cloud)}}} \\
Bayesian & $0.56^{+0.08}_{-0.09}$ & $1.25^{+0.12}_{-0.12}$ & $-5.11^{+0.17}_{-0.19}$& & & & & 3.98 & 572.69\\ 
[0.2cm]
ML & \tr{$0.60^{+0.00}_{-0.01}$} & \tr{$1.22^{+0.05}_{-0.00}$} & \tr{$-4.98^{+0.04}_{-0.09}$}& & & & & \tr{3.87} & & \tr{1.26} & 0.970 \\
[0.2cm]
\hline
& \multicolumn{6}{c}{\textit{{Two-Stream Approximation (Non-Uniform Cloud)}}} \\
Bayesian & $0.42^{+0.17}_{-0.13}$ & $1.24^{+0.17}_{-0.14}$ & $-4.85^{+0.28}_{-0.29}$& & & & & 2.43 & 622.71\\ 
[0.2cm]
ML & \tr{$0.49^{+0.01}_{-0.05}$} & \tr{$1.26^{+0.06}_{-0.03}$} & \tr{$-4.99^{+0.03}_{-0.03}$}& & & & & \tr{1.22} & & \tr{0.313} & 0.968 \\
[0.2cm]
\hline
\hline
& \multicolumn{7}{c}{{\textbf{Equilibrium Offset}}} \\
\hline

\multirow{2}{*}{} & \multicolumn{1}{c}{C/O} & \multicolumn{1}{c}{log[M/H]} & \multicolumn{1}{c}{log(SO$_2$)} & \multicolumn{1}{c}{$\delta(\text{H}_2\text{O})$} & \multicolumn{1}{c}{$\delta(\text{C}\text{O}_2)$} & \multicolumn{1}{c}{$\delta(\text{C}\text{O})$} & \multicolumn{1}{c}{$\delta(\text{C}\text{H}_4)$} & \\
\bottomrule

& \multicolumn{6}{c}{\textit{{One-Stream Approximation (Clear)}}} \\
Bayesian & $0.29^{+0.13}_{-0.06}$ & $0.80^{+0.17}_{-0.13}$ & $-5.74^{+0.19}_{-0.18}$ & $0.15^{+0.11}_{-0.06}$ & $1.35^{+0.44}_{-0.56}$ & $1.31^{+0.43}_{-0.46}$ & $0.02^{+0.02}_{-0.01}$ & 3.59 & 584.28\\
[0.2cm]
ML & \tr{$0.23^{+0.06}_{-0.03}$} & \tr{$0.70^{+0.09}_{-0.00}$} & \tr{$-5.39^{+0.22}_{-0.36}$} & \tr{$0.11^{+0.03}_{-0.01}$} & \tr{$1.01^{+0.44}_{-0.01}$} & \tr{$1.27^{+0.09}_{-0.08}$} & \tr{$0.03^{+0.02}_{-0.02}$} & \tr{3.50} & & \tr{1.15}& 0.923\\
[0.2cm]
\hline
& \multicolumn{6}{c}{\textit{{Two-Stream Approximation (Uniform Cloud)}}} \\
Bayesian & $0.32^{+0.14}_{-0.08}$ & $0.82^{+0.19}_{-0.15}$ & $-5.71^{+0.20}_{-0.21}$ & $0.19^{+0.13}_{-0.07}$ & $1.32^{+0.43}_{-0.52}$ & $1.35^{+0.41}_{-0.43}$ & $0.02^{+0.02}_{-0.01}$ & 3.91 & 583.84\\
[0.2cm]
ML & \tr{$0.37^{+0.12}_{-0.03}$} & \tr{$0.91^{+0.06}_{-0.06}$} & \tr{$-5.67^{+0.65}_{-0.31}$} & \tr{$0.18^{+0.01}_{-0.09}$} & \tr{$1.43^{+0.14}_{-0.47}$} & \tr{$1.29^{+0.10}_{-0.09}$} & \tr{$0.05^{+0.00}_{-0.00}$} & \tr{2.75} & & \tr{0.940} & 0.950\\
[0.2cm]
\hline
& \multicolumn{6}{c}{\textit{{Two-Stream Approximation (Non-Uniform Cloud)}}} \\
Bayesian & $0.44^{+0.16}_{-0.14}$ & $1.28^{+0.28}_{-0.24}$ &  $-4.76^{+0.41}_{-0.37}$ & $1.30^{+0.42}_{-0.43}$ & $1.27^{+0.41}_{-0.46}$ & $0.81^{+0.67}_{-0.54}$ & $0.99^{+0.57}_{-0.56}$ & 2.57 & 621.53\\
[0.2cm]
ML & \tr{$0.48^{+0.01}_{-0.04}$} & \tr{$1.21^{+0.06}_{-0.01}$} & \tr{$-4.03^{+0.02}_{-0.96}$} & \tr{$1.05^{+0.28}_{-0.03}$} & \tr{$1.18^{+0.34}_{-0.12}$} & \tr{$0.82^{+0.00}_{-0.00}$} & \tr{$0.95^{+0.01}_{-0.01}$} & \tr{2.75} & & \tr{1.07} & 0.935 \\
[0.2cm]
\hline
\end{tabular}
}
\label{Table:all_params}
\end{table*}


\begin{figure*} []\label{fig:vmr}
    \centering
    \begin{minipage}{0.49\textwidth}
        \centering
        \includegraphics[width=\textwidth]{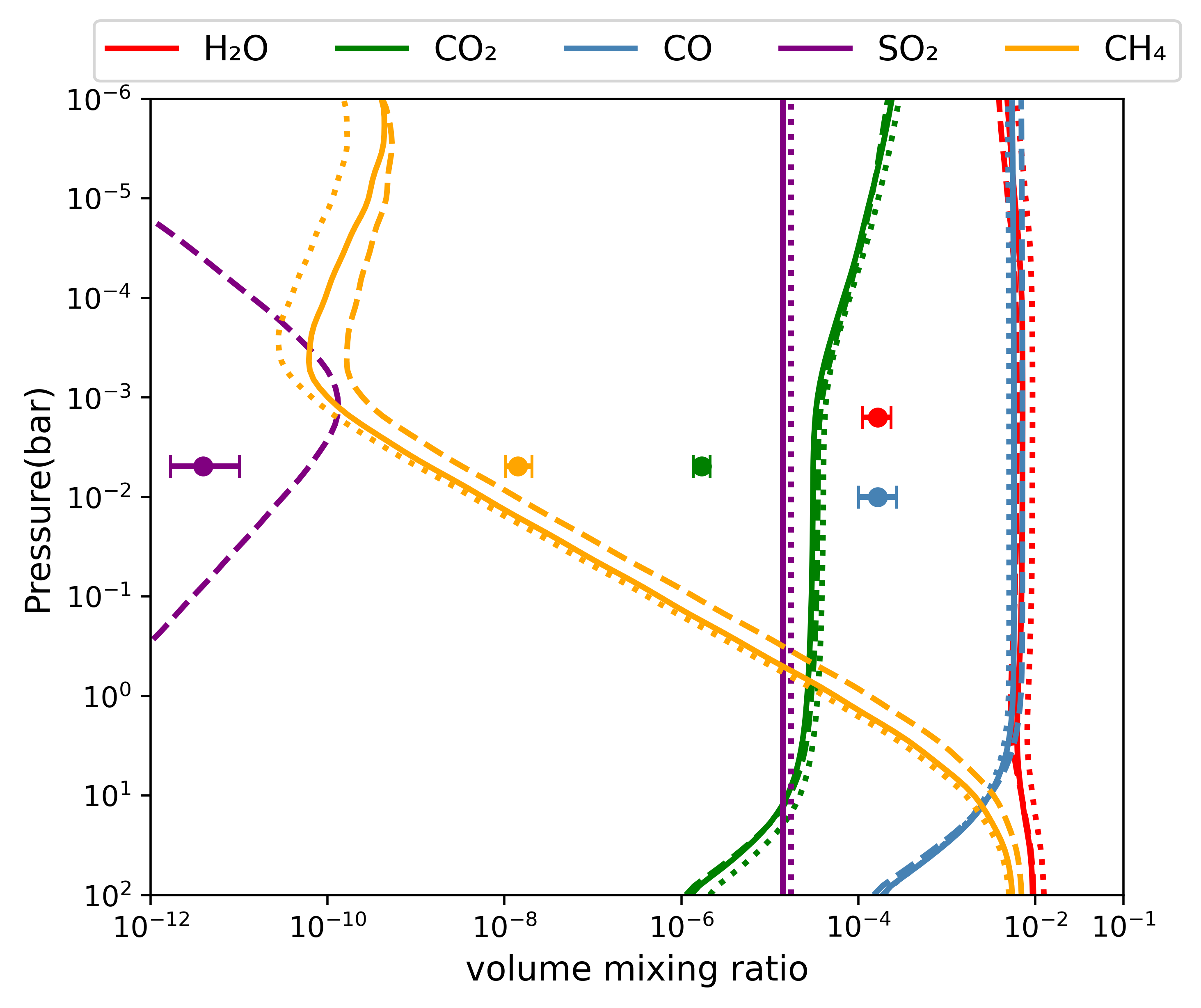}
        \parbox{0.8\linewidth}{\centering (a)}
        \label{fig:vmr_free}
        
    \end{minipage}
    \hfill
    \begin{minipage}{0.48\textwidth}
        \centering
        \vspace{0.6cm}
        \includegraphics[width=\textwidth]{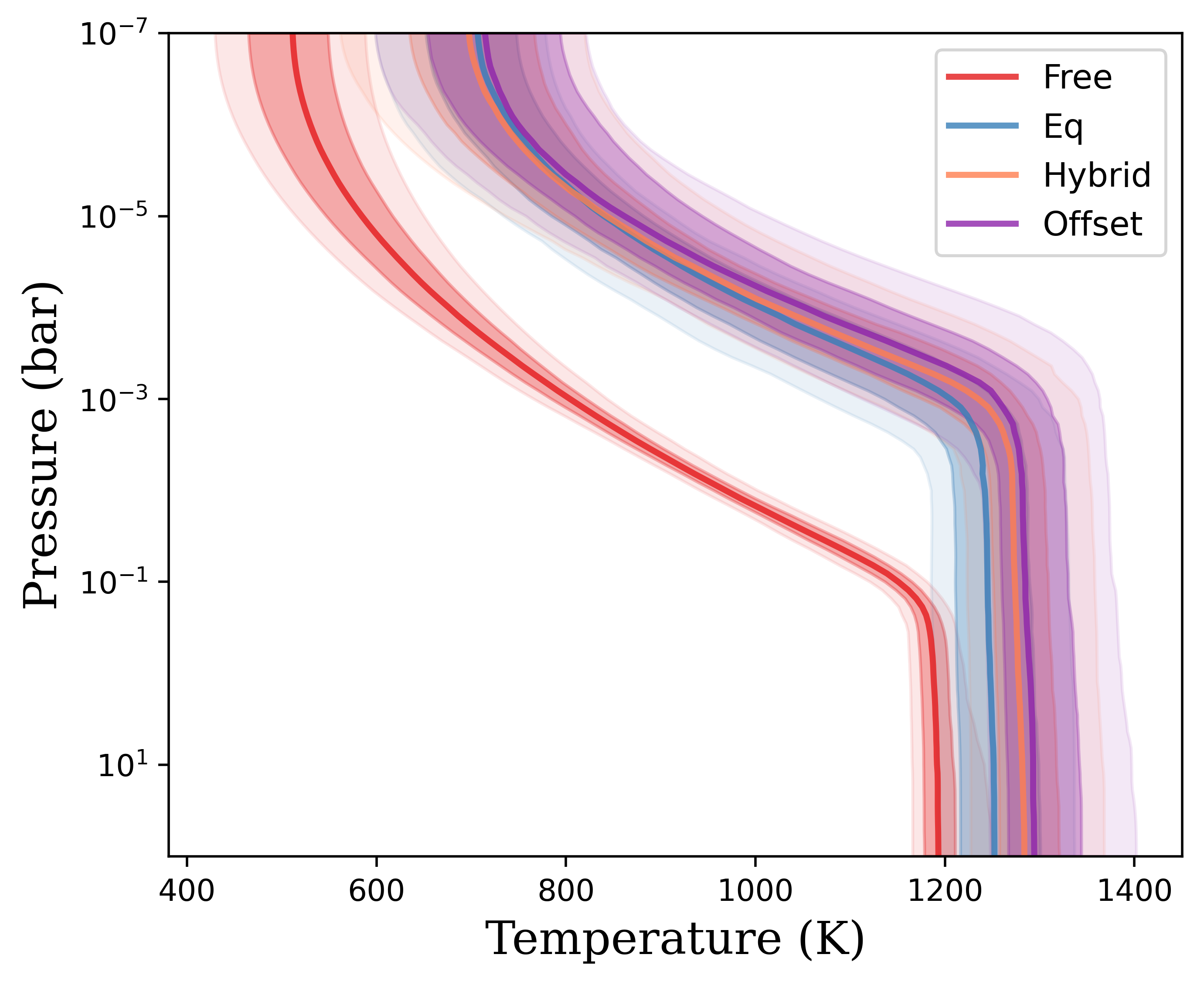}
        \parbox{0.8\linewidth}{\centering (b) }
        \label{fig:pt_free}
    \end{minipage}
    
    \caption{(a) Retrieved Volume Mixing Ratio (VMR) profiles for key molecular species in the atmosphere of WASP-69b under different chemical modeling assumptions and the presence of patchy non-uniform MgSiO$_3$ aerosol, using Bayesian retrieval: free chemistry (points), equilibrium chemsitry (dashed), hybrid equilibrium chemistry(solid), and equilibrium offset chemistry (dotted). 
    (b) Retrieved P-T profile for non-uniform aerosol models and all chemistries using Bayesian retrieval.}
\label{fig:vmr_pt}
\end{figure*}

\begin{figure*} []\label{fig:vmr}
    \centering
    \begin{minipage}{0.49\textwidth}
        \centering
        \includegraphics[width=\textwidth]{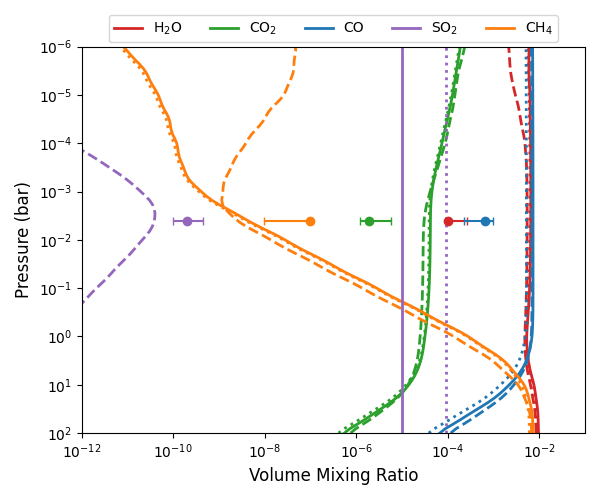}
        \parbox{0.8\linewidth}{\centering (a)}
        \label{fig:vmrML_free}
        
    \end{minipage}
    \hfill
    \begin{minipage}{0.48\textwidth}
        \centering
        \vspace{0.6cm}
        \includegraphics[width=\textwidth]{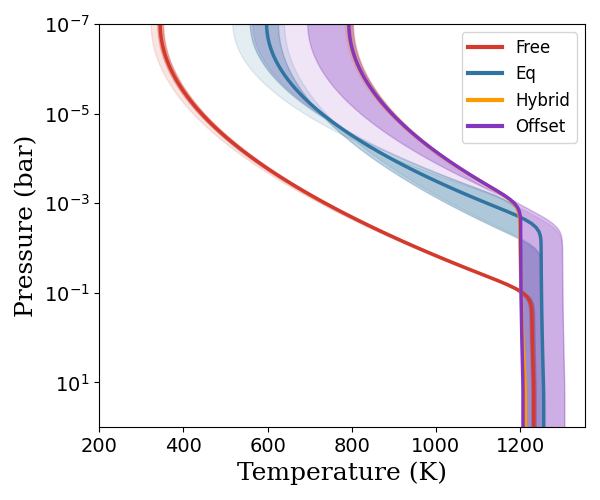}
        \parbox{0.8\linewidth}{\centering (b) }
        \label{fig:pt_free_ML}
    \end{minipage}
    
    \caption{\tr{(a) Retrieved Volume Mixing Ratio (VMR) profiles for key molecular species in the atmosphere of WASP-69b under different chemical modeling assumptions and the presence of patchy non-uniform MgSiO$_3$ aerosol, using Machine leaning (ML) retrieval: free chemistry (points), equilibrium chemsitry (dashed), hybrid equilibrium chemistry(solid), and equilibrium offset chemistry (dotted). (b) Retrieved P-T profile for non-uniform aerosol models and all chemistries using ML retrieval.}}  
\label{fig:ML_vmr_pt}
\end{figure*}


\section{\textbf{RESULTS}} \label{results}

We now apply \texttt{NEXOTRANS} to retrieve the dayside atmospheric properties of the full emission spectrum of WASP-69b, observed with JWST \citep{schlawin2024multiple}. The 2--5 $\mu$m observations were obtained using NIRCam and reduced using the \texttt{Eureka!} pipeline, whereas the 5--12 $\mu$m observations were taken with MIRI and reduced using the \texttt{Tshirt} pipeline. We downloaded the reduced spectra directly from the atmospheric spectroscopy tables of the NASA Exoplanet Archive \footnote{\url{https://exoplanetarchive.ipac.caltech.edu/cgi-bin/atmospheres/nph-firefly?atmospheres}}.

WASP-69b is a hot Saturn-mass exoplanet ($0.29\,M_J$) with a radius 1.11 times that of Jupiter, a surface gravity of 5.8 m/s$^2$, and an equilibrium temperature of 988 K. It orbits a K-type star with an effective temperature of $\sim4700$ K and a radius 0.86 times that of the Sun.
  
To investigate the atmospheric composition of WASP-69b, we employ four distinct chemistry models, as detailed in Section~\ref{subsubsection : atmospheric profiles}. We begin with the simplistic free-chemistry approach, which does not impose any constraints from thermochemical equilibrium, allowing each molecular abundance to vary independently. Next, we consider the assumption of chemical equilibrium, where molecular abundances are determined solely by local temperature and pressure. Based on the possibility of dayside photochemistry and recent inferences of SO$_2$ in WASP-69b's atmosphere \citep{schlawin2024multiple}, we also explore two approximate disequilibrium chemistry models: the modified hybrid equilibrium and the modified equilibrium-offset chemistry frameworks.
Given that SO$_2$ is likely produced through photochemical processes, such as photodissociation driven by stellar irradiation on the dayside of the planet, we apply these disequilibrium approaches using the \texttt{NEXOTRANS} retrieval framework. These methods relax the assumptions of strict equilibrium and enable a proxy treatment of disequilibrium processes, including photochemistry, which are expected to shape the observable composition of WASP-69b’s atmosphere.

We conduct retrievals using both one- and two-stream radiative transfer approximations, considering a clear atmosphere and another including aerosols (MgSiO$_3$) for the respective approximations. Based on prior findings in \citet{schlawin2024multiple}, we include the key molecular species H$_2$O, CO$_2$, CO, SO$_2$, and CH$_4$ in our analysis.  

The Bayesian retrievals are performed at a model resolution of 15,000 and 1000 live points to balance computational efficiency and accuracy in our retrievals. The free parameters for the various models explored are listed in Table \ref{Table:free_parameter}. Retrievals were also performed with machine learning (ML) algorithm using \texttt{Stacking Regressor}. Models were trained using  \tr{1,00,000} simulated spectra from the \texttt{NEXOTRANS} forward model separately for free chemistry, equilibrium chemistry, hybrid and equilibrium offset chemistry, for both one- and two-stream approximations. The retrieved values are shown in Table \ref{Table:all_params}.

\subsection{\textbf{Retrieved Abundances}}


The \texttt{NEXOTRANS} retrievals statistically favor an atmospheric model that includes aerosols, in agreement with the findings of \citet{schlawin2024multiple}. The retrieved molecular abundances for both the PyMultiNest-based Bayesian retrievals and the machine learning retrievals using the Stacking Regressor are summarized in Table~\ref{Table:all_params} for the different atmospheric models considered.

The overall absorption on the emission spectra is shaped by H$_2$O, with clearly visible absorption by CO$_2$ at 4.3 $\mu$m along with a minor contribution of CO redward of it, contributing to the low flux at 4.6 $\mu$m. The possible absorption due to SO$_2$ is seen in the free, hybrid equilibrium and equilibrium offset chemistry retrievals at 7-8 $\mu$m except for the case of equilibrium and non-uniform aerosol + free chemistry models. A similar potential inference for SO$_2$ in this wavelength region was previously reported by \citet{schlawin2024multiple}. The equilibrium model with non-uniform aerosols has a reduced $\chi^2$ value of 2.75 as compared to those of the hybrid and equilibrium offset with values 2.43 and 2.57 respectively. Clearly, the retrievals favor an atmospheric chemical condition that is not in equilibrium. This suggests the possible formation of SO$_2$ through photochemistry in WASP-69 b's dayside atmosphere. No obvious CH$_4$ features are observed
in the emission spectrum. The high temperatures ($\geq$1000 K at 10$^{-3}$--10$^{-4}$ bar) inferred from the retrievals result in low CH$_4$ abundance, with H$_2$O absorption dominating and obscuring any CH$_4$ features.

\tr{The volume mixing ratio (VMR) profiles retrieved using the Bayesian method for various molecular species under free (points), equilibrium (dashed), hybrid equilibrium (solid), and equilibrium offset (dotted) chemistry with non-uniform aerosols are shown in Figure~\ref{fig:vmr_pt}(a). The VMR profiles retrieved using the machine learning algorithm are shown in Figure~\ref{fig:ML_vmr_pt}(a). The broader and more asymmetric error bars in the ML results, compared to those from the Bayesian method, reflect the empirical nature of the uncertainty estimation, which depends on the spread of the prediction variance across the posterior samples, in contrast to the rigorous error propagation in the Bayesian approach.} Apart from free chemistry, across all other models, H$_2$O, CO, CO$_2$, and CH$_4$ show only minor deviations from their equilibrium profiles, indicating that thermochemical equilibrium largely governs their distribution. H$_2$O, CO$_2$ and CO remain the dominant species across all models in the photospheric region of the atmosphere. CH$_4$ is depleted in the upper atmosphere but shows an increasing trend at higher pressures, consistent with expectations for a moderately irradiated warm exoplanet atmosphere.
The most significant deviations across models occur for SO$_2$, where it has an enhanced abundance in the hybrid and offset equilibrium cases as compared to the equilibrium or free assumptions where the VMR ranges between 10$^{-10}$ and 10$^{-14}$. VMR values between 10$^{-4}$ and 10$^{-6}$ in the hybrid and offset models suggest the presence of disequilibrium mechanisms such as photochemistry. 

As seen in the case of the equilibrium offset retrievals (see Table~\ref{Table:all_params}), the assumption of cloud coverage--whether uniform or non-uniform (patchy)--has a notable impact on the retrieved molecular abundances. For uniform clouds, the retrieved H$_2$O volume mixing ratio is slightly depleted relative to the equilibrium abundance, with a multiplicative offset of $0.19^{+0.13}_{-0.07}$. In contrast, the non-uniform aerosol case yields a higher value of $1.30^{+0.42}_{-0.37}$, indicating a slight increase from equilibrium. The most striking difference is observed for CH$_4$, where the non-uniform aerosol case shows an offset factor of $0.99^{+0.57}_{-0.56}$ hinting no significant depletion, whereas the uniform cloud model results in a strong depletion with an offset of $0.02^{+0.02}_{-0.01}$. While this depletion may appear significant, it still largely follows the shape of the equilibrium VMR profile, as shown in Figure~\ref{fig:vmr_pt}(a). \tr{Additionally, as evident from the ML profiles in Figure~\ref{fig:ML_vmr_pt}(a), CH$_4$ appears to be highly sensitive to even small variations in the P–T profiles, resulting in discrepancies between the Bayesian and machine learning models in the CH$_4$ VMR profiles of the upper atmosphere for the hybrid and equilibrium offset chemistry cases.} Although the retrievals statistically favor a non-uniform aerosol assumption, it is important to adopt a physically motivated framework for aerosol formation and distribution in order to accurately model their influence on the observed spectrum.


\begin{table*}
\centering
\caption{Corrected retrieved aerosol properties for uniform and non-uniform treatments across different chemistry models, comparing Bayesian and Machine Learning (ML) retrievals.}
\resizebox{0.85\textwidth}{!}{
\begin{tabular}{llcccccc}
\toprule
\hline
Treatment & Retrieval & $\log(X_{\text{MgSiO}_3})$ & $\log(r_{\text{MgSiO}_3})$ & $\log(P_{\text{MgSiO}_3, \text{deck}})$ & $f_{\text{MgSiO}_3}$ & $\phi_{\text{MgSiO}_3}$ & Red.$\chi^2$ \\
\midrule
\multicolumn{8}{c}{\textit{\textbf{Free Chemistry}}} \\
\multirow{2}{*}{Uniform} & Bayesian & $-10.37^{+6.18}_{-6.28}$ & $-1.05^{+1.35}_{-1.31}$ & $-1.91^{+2.60}_{-2.65}$ & $0.49^{+0.27}_{-0.26}$ & -- & 3.51\\
& ML & \tr{$-10.12^{+0.37}_{-0.38}$} & \tr{$-0.15^{+0.05}_{-0.06}$} & \tr{$-1.94^{+0.44}_{-0.46}$} & \tr{$0.47^{+0.02}_{-0.01}$} & -- & \tr{2.29} \\[0.2cm]
\multirow{2}{*}{Non-Uniform} & Bayesian & $-9.51^{+6.64}_{-6.58}$ & $0.00^{+0.64}_{-0.85}$ & $-1.80^{+2.34}_{-2.58}$ & $0.45^{+0.29}_{-0.23}$ & $0.21^{+0.03}_{-0.03}$ & 2.37\\
& ML & \tr{$-9.87^{+0.09}_{-0.08}$} & \tr{$-0.18^{+0.00}_{-0.02}$} & \tr{$-2.34^{+0.00}_{-0.00}$} & \tr{$0.41^{+0.01}_{-0.00}$} & \tr{$0.23^{+0.01}_{-0.01}$} & \tr{1.40} \\
[0.2cm]
\hline
\hline
\multicolumn{8}{c}{\textit{\textbf{Equilibrium}}} \\
\multirow{2}{*}{Uniform} & Bayesian & $-10.07^{+5.62}_{-5.92}$ & $-0.90^{+1.20}_{-1.28}$ & $-1.90^{+2.41}_{-2.51}$ & $0.49^{+0.25}_{-0.23}$ & -- & 4.63\\
& ML & \tr{$-9.91^{+0.83}_{-0.05}$} & \tr{$-0.87^{+0.06}_{-0.02}$} & \tr{$-1.02^{+0.02}_{-0.98}$} & \tr{$0.33^{+0.16}_{-0.13}$} & -- & \tr{3.56} \\[0.2cm]
\multirow{2}{*}{Non-Uniform} & Bayesian & $-3.30^{+1.27}_{-1.18}$ & $0.75^{+0.18}_{-0.77}$ & $-2.63^{+2.66}_{-2.11}$ & $0.29^{+0.27}_{-0.14}$ & $0.43^{+0.07}_{-0.05}$ & 2.75\\
& ML & \tr{$-3.31^{+0.09}_{-0.03}$} & \tr{$-0.61^{+0.01}_{-0.00}$} & \tr{$-2.53^{+0.02}_{-0.93}$} & \tr{$0.11^{+0.15}_{-0.00}$} & \tr{$0.60^{+0.00}_{-0.00}$} & \tr{2.71} \\
[0.2cm]
\hline
\hline
\multicolumn{8}{c}{\textit{\textbf{Hybrid Equilibrium}}} \\
\multirow{2}{*}{Uniform} & Bayesian & $-10.62^{+5.97}_{-6.12}$ & $-1.03^{+1.29}_{-1.25}$ & $-1.91^{+2.46}_{-2.55}$ & $0.51^{+0.25}_{-0.26}$ & -- & 3.98\\
& ML & \tr{$-10.69^{+0.17}_{-0.17}$} & \tr{$-1.96^{+0.80}_{-0.03}$} & \tr{$-2.14^{+0.08}_{-0.07}$} & \tr{$0.52^{+0.01}_{-0.00}$} & -- & \tr{3.87} \\[0.2cm]
\multirow{2}{*}{Non-Uniform} & Bayesian & $-3.18^{+1.27}_{-1.44}$ & $-0.23^{+0.09}_{-0.07}$ & $-2.52^{+2.55}_{-2.12}$ & $0.34^{+0.22}_{-0.16}$ & $0.46^{+0.04}_{-0.04}$ & 2.43\\
& ML & \tr{$-3.23^{+0.05}_{-0.04}$} & \tr{$-0.23^{+0.02}_{-0.03}$} & \tr{$-2.45^{+0.07}_{-0.05}$} & \tr{$0.34^{+0.02}_{-0.02}$} & \tr{$0.45^{+0.01}_{-0.01}$} & \tr{1.22} \\
[0.2cm]
\hline
\hline
\multicolumn{8}{c}{\textit{\textbf{Equilibrium Offset}}} \\
\multirow{2}{*}{Uniform} & Bayesian & $-10.24^{+5.84}_{-6.00}$ & $-0.93^{+1.22}_{-1.27}$ & $-2.04^{+2.57}_{-2.48}$ & $0.50^{+0.26}_{-0.25}$ & -- & 3.91\\
& ML & \tr{$-10.52^{+0.42}_{-0.38}$} & \tr{$-0.97^{+0.02}_{-0.02}$} & \tr{$-2.02^{+0.09}_{-0.19}$} & \tr{$0.47^{+0.10}_{-0.07}$} & -- & \tr{2.75} \\[0.2cm]
\multirow{2}{*}{Non-Uniform} & Bayesian & $-3.52^{+1.40}_{-8.10}$ & $-0.22^{+0.10}_{-0.08}$ & $-2.18^{+2.28}_{-2.10}$ & $0.36^{+0.24}_{-0.17}$ & $0.48^{+0.09}_{-0.04}$ & 2.57\\
& ML & \tr{$-3.76^{+0.04}_{-0.04}$} & \tr{$-0.26^{+0.04}_{-0.02}$} & \tr{$-2.05^{+0.07}_{-0.09}$} & \tr{$0.34^{+0.03}_{-0.02}$} & \tr{$0.49^{+0.01}_{-0.07}$} & \tr{2.75} \\
\bottomrule
\end{tabular}
}
\label{tab:cloud_params_ML_corrected}
\end{table*}

We also retrieve an overall super-solar metallicity ranging from log[M/H] = 0.80 to 1.27 under a clear atmospheric model (Table \ref{Table:all_params}), indicating an enrichment in heavy elements relative to solar. The retrieved C/O ratio spans a broad range, from subsolar to moderately solar (0.29 -- 0.56), suggesting diverse carbon-oxygen chemistry pathways. Models incorporating uniform and non-uniform aerosol contributions also follow this trend, with C/O varying between 0.30 and 0.56, and log[M/H] between 0.82 and 1.28. The molecular and elemental abundances retrieved from Stacking Regressor (machine learning) retrievals also show consistent results as compared to its Bayesian counterpart with a maximum median super-solar C/O value of \tr{$0.60$} and a maximum super-solar metallicity of \tr{$1.26$}. 

This metallicity enhancement aligns with the trends observed for other warm gas giants, where lower-mass planets tend to show higher metal enrichment. The presence of aerosols further complicates this interpretation, as different models suggest variations in cloud coverage and chemical interactions. Overall, these findings indicate that WASP-69b's atmosphere is enriched in heavy elements and possibly influenced by photochemical processes, vertical mixing etc.

\subsection{\textbf{Retrieved Thermal Profile}}


\tr{The retrieved temperature–pressure (T–P) profiles for different chemical models incorporating non-uniform aerosol distributions are shown in Figure~\ref{fig:vmr_pt}(b) and Figure~\ref{fig:ML_vmr_pt}(b), corresponding to the Bayesian and ML methods, respectively.}. The retrievals impose precise constraints on the thermal structure of the dayside atmosphere of WASP-69b, revealing a temperature gradient that ranges from approximately 500 K in the upper atmosphere to 1400 K at deeper pressures. The profiles show a smooth, monotonic decrease in temperature with altitude, consistent with expectations for hot Jupiters that lack significant stratospheric heating \citep{Zahnle_2009}.

The retrieved photospheric temperature, corresponding to the effective emission layer of the dayside spectrum, is approximately between 900 -- 1000 K. This result closely aligns with the planet’s equilibrium temperature, $T_{\text{eq}} = 988$ K, under the assumption of efficient radiative redistribution and negligible Bond albedo. Furthermore, the retrieved thermal profile shows an approximately isothermal structure at pressures deeper than the photospheric level, a characteristic feature commonly observed in highly irradiated hot Jupiter. This quasi-isothermal region arises due to strong radiative absorption in the upper layers, which thermally decouples the deeper atmosphere from stellar irradiation, resulting in a nearly constant temperature structure at high optical depths \citep{spiegel2013thermal, guillot2010radiative}.

\tr{In the equilibrium, hybrid, and offset chemistry models, the abundances and temperature profiles are linked through the equilibrium chemistry grid. This coupling causes the VMR and T–P profiles to co-vary, leading these models to naturally converge toward a similar family of T–P structures. In contrast, in the free chemistry case, we assume vertically constant VMRs, and since the chemistry does not dictate the vertical dependence of the abundances, the retrieval compensates by adjusting the temperature. Consequently, due to this flexibility between abundances and temperatures, the retrieved T–P profiles for the free chemistry case differ from those of the other models, as shown in Figure~\ref{fig:vmr_pt}(b). Although the T–P profile retrieved using the ML method closely matches the Bayesian profile, some differences remain. These differences arise from subtle variations in the retrieved T–P parameters from the ML method compared to the Bayesian retrieval, such as $\alpha_1$, which controls the slope of the profile.}


Overall, the retrieved T–P profile of WASP-69b is consistent with an atmosphere in radiative equilibrium, lacking evidence for a thermal inversion. This supports a scenario in which molecular absorption and large-scale atmospheric dynamics govern the observed thermal emission properties of the planet. The temperature structure retrieved using \texttt{NEXOTRANS} closely aligns with the earlier results reported by \citet{schlawin2024multiple} and any differences in the shape of the profiles emerges due to the P-T parametrization used (e.g., Guillot vs Madhusudhan profile).

\begin{figure*}[]
    \centering
    \includegraphics[width=0.90\linewidth]{
    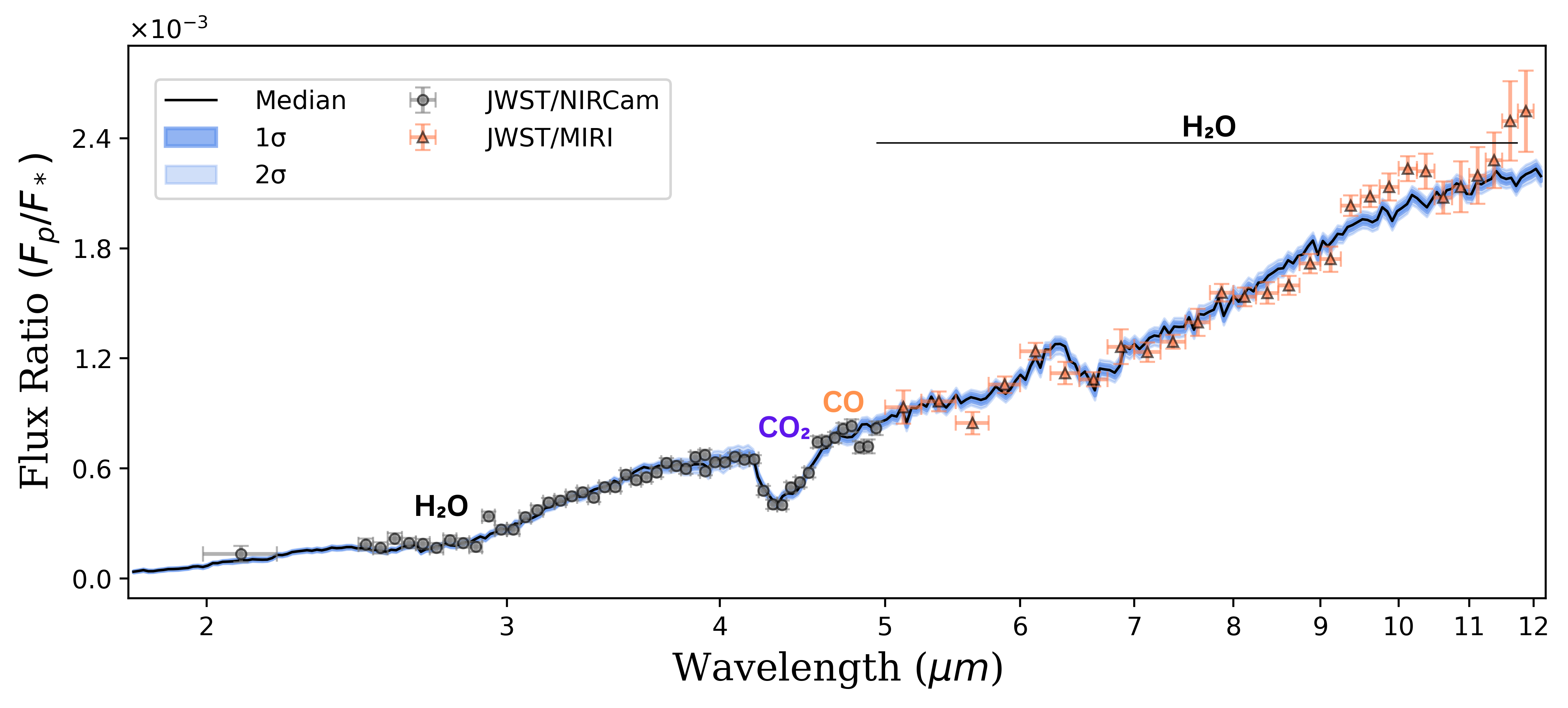}
    \parbox{0.8\linewidth}{\centering (a) Overall Bayesian best-fit spectrum with free chemistry and non-uniform aerosol. }
    \label{fig:spectrum}
    \hspace{-0.13cm}
    \includegraphics[width=0.90\linewidth]{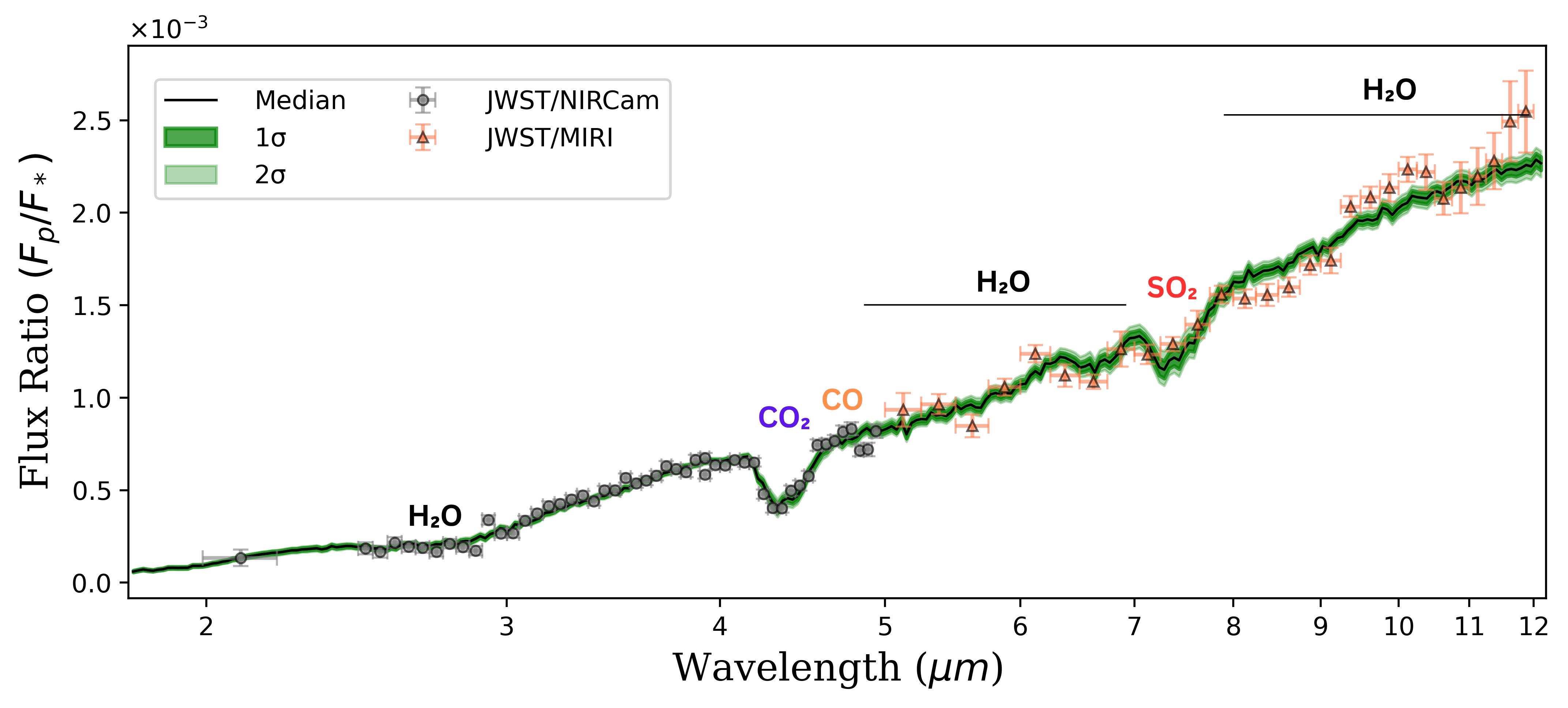}
    \parbox{0.8\linewidth}{\centering (b) Best-fit spectrum utilizing the NEXOCHEM chemistry grid. This corresponds to the hybrid equilibrium chemistry model with non-uniform aerosol.}
    \label{fig:hybrid_ML_spectra}
    \caption{Figure (a) shows the best-fit spectrum among all the retrievals performed, belonging to the model assuming free chemistry with non-uniform aerosol. It is clear that the overall best-fit model doesn't have the SO$_2$ absorption feature around 7-8 $\mu$m as compared to the other. Figure (b) 
     shows the best-fit spectrum among the models utilizing the NEXOCHEM chemistry grid, the spectrum shown is for the model assuming hybrid chemistry with non-uniform aerosol configuration. It clearly has the absorption dip due to SO$_2$.}
    \label{best-fit spectrum}
\end{figure*}
\subsection{\textbf{Retrieved Aerosol Parameters}}

The retrieved aerosol (cloud) properties of WASP-69b reveal key differences in their abundance, size and coverage depending on the adopted chemistry model and coverage treatment (Table \ref{tab:cloud_params_ML_corrected}). The retrieved MgSiO$_3$ log(VMR) ($X_{\text{MgSiO}_3}$) varies significantly, with non-uniform aerosol and equilibrium-based models favoring higher values than other models. Notably, non-uniform treatments also consistently retrieve larger particles, whereas uniform treatments favor smaller particles. \tr{The aerosol parameters retrieved using the ML method are also consistent with the Bayesian results.} This trend suggests that considering spatial variations in aerosol coverage allow for larger condensates.
The retrieved opaque cloud deck base pressure log(P$_{\text{MgSiO}_3, \text{deck}}$) is also model-dependent, with non-uniform cases mostly favoring lower pressures, indicating that spatially variable cloud is formed at higher altitudes. By contrast, uniform cloud models comparatively place the cloud base at higher pressures, implying deeper decks. The presence of high-altitude aerosol decks was also retrieved in \citet{schlawin2024multiple}, located between 10$^{-4.5}$ and 10$^{-6}$ bar, significantly above the expected condensation level of MgSiO$_3$, which lies near 10 bar in WASP-69b’s atmosphere based on its temperature–pressure profile. \citet{schlawin2024multiple} claims that an extreme lofting mechanism \citep{Charnay_2015} is required to sustain silicate clouds at such low pressures. Our retrieval results independently arrive at the same conclusion, further supporting the hypothesis of vertically extended aerosols in the WASP-69b's atmosphere.

The retrieved aerosol slope parameter ($f_{\text{MgSiO}_3}$) shows moderate variation across models but all values within 1$\sigma$ of each other. The values mostly indicate a moderately steady decrease in the aerosol VMR profiles. Similarly, the cloud coverage fraction ($\mathrm{\phi}_{\mathrm{MgSiO}_3}$
) is consistent around 0.43--0.48 in non-uniform cloud models except free chemistry, implying that nearly half of the planetary atmosphere is covered by clouds and this fraction remains stable across equilibrium based chemistry models. 
Importantly, the reduced $\chi^2$ values indicate that non-uniform cloud models provide a statistically better fit across all chemistry models, reinforcing the necessity of accounting for spatial heterogeneity of clouds in retrievals.

These results indicate that incorporating additional flexibility in chemistry and spatial cloud distribution can significantly improve retrieval fits. The preference for larger aerosol particles in non-uniform cloud models suggests that cloud growth and aggregation processes may be more efficient in atmospheres exhibiting spatial variability in cloud coverage. Moreover, the retrieval of higher-altitude clouds in these models may point to enhanced vertical mixing or the presence of photochemically produced hazes. Future work involving three-dimensional cloud modeling and high-resolution, phase-resolved observations-coupled with self-consistent treatments of cloud microphysics will be crucial in constraining the spatial and temporal evolution of clouds in warm exoplanetary atmospheres.

\section{\textbf{Discussions}} \label{discussion}

In this section, we discuss the statistically and chemically best-fit model and the implications the results provide on the atmosphere of WASP-69b.

\begin{figure*}[]
    \centering
    \includegraphics[width=1.0\linewidth]{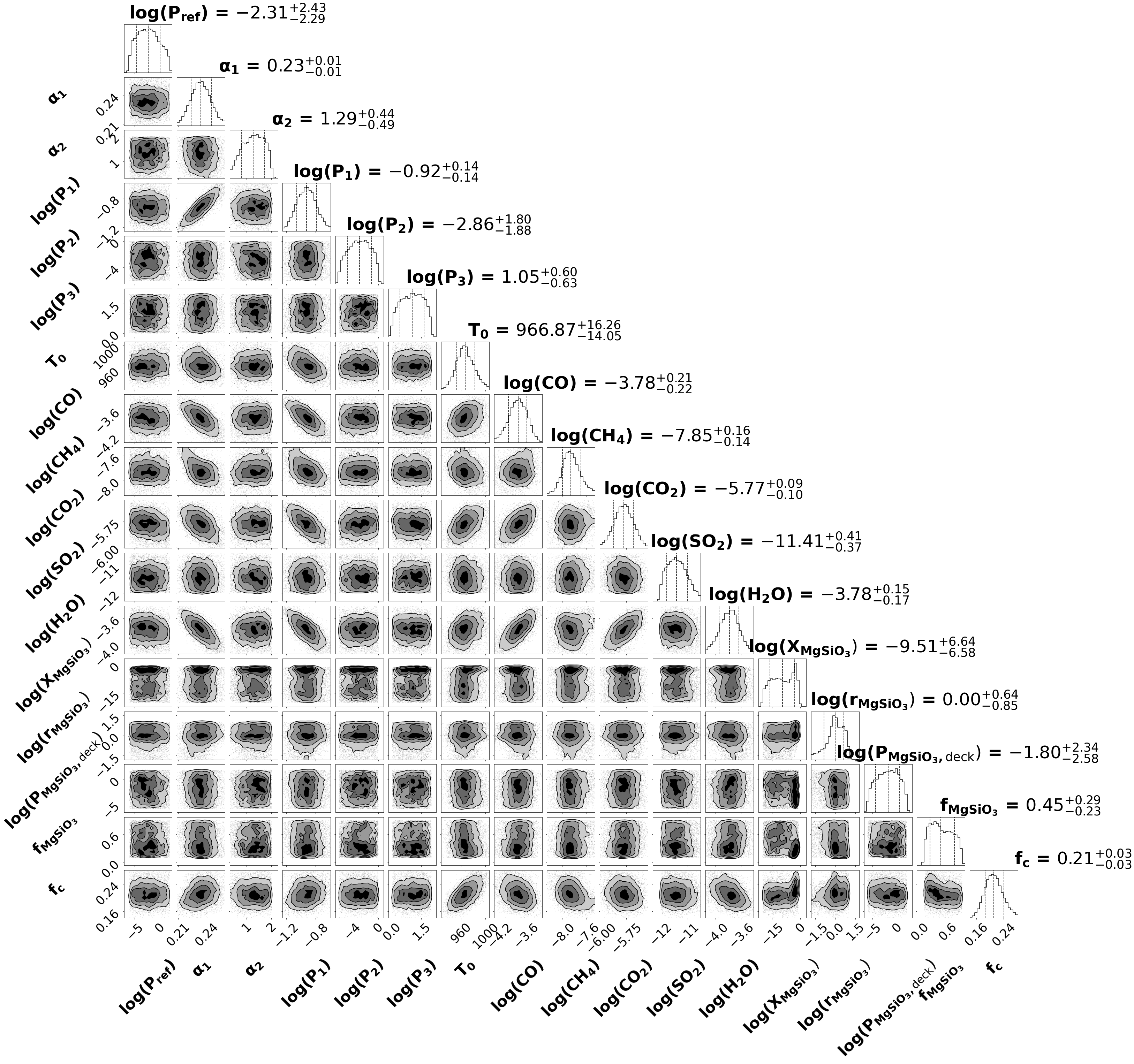}
    \parbox{0.8\linewidth}{}
    \caption{Full posterior distribution for the best-fit non-uniform Mie-scattering MgSiO$_3$ aerosol model, assuming free chemistry (with PyMultiNest). The corner plot shows the correlations between pairs of retrieved parameters and the marginalized distributions for each parameter. The retrieved median values and corresponding 1$\sigma$ uncertainties are also indicated.}
    \label{fig:main_corner}
\end{figure*}

\begin{figure*}[]
    \centering
    \includegraphics[width=1.0\linewidth]{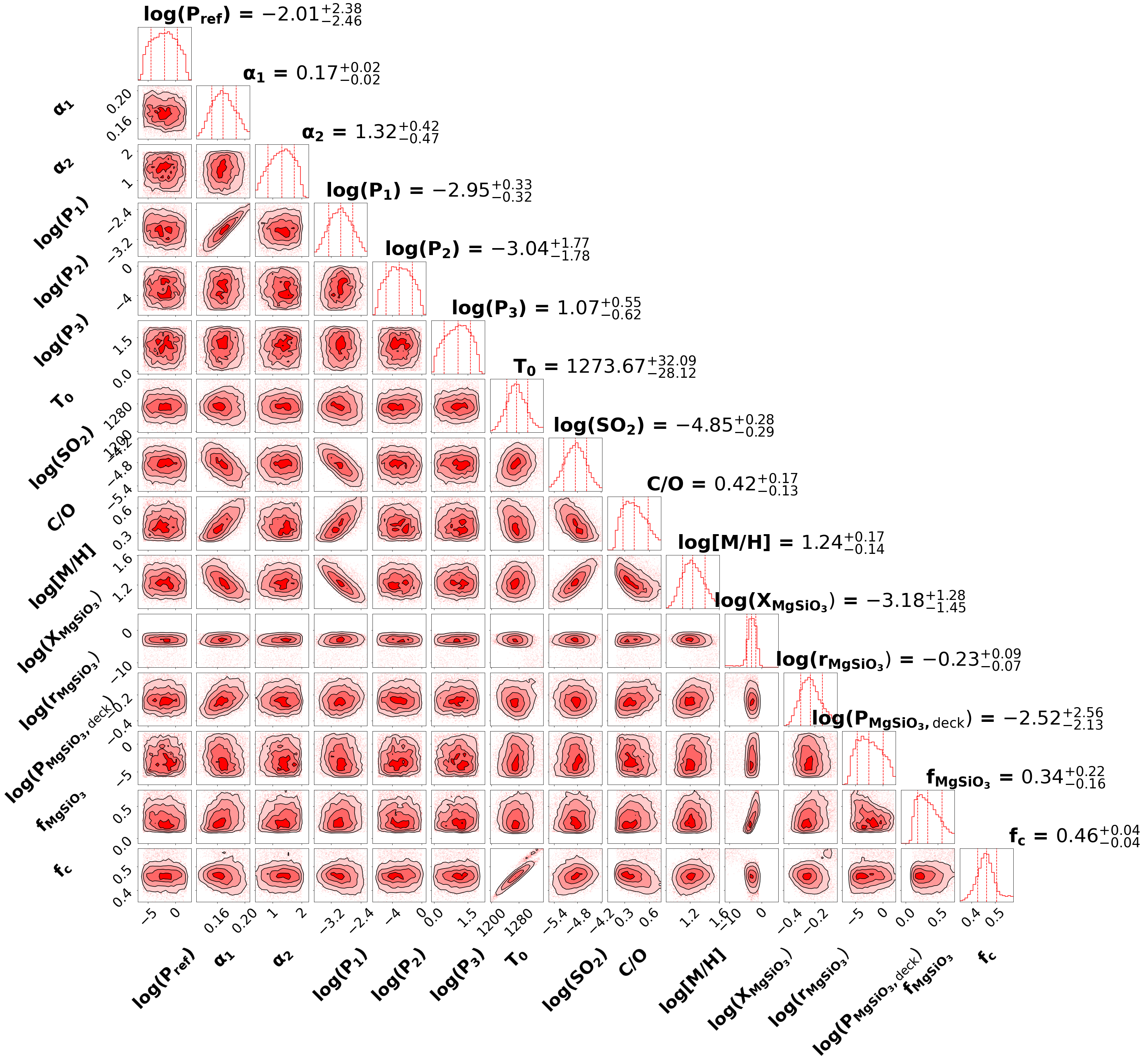}
    \parbox{0.8\linewidth}{}
    \caption{Full posterior distribution for the second best-fit non-uniform Mie-scattering MgSiO$_3$ aerosol model, assuming hybrid equilibrium chemistry (with PyMultiNest). The corner plot shows the correlations between pairs of retrieved parameters and the marginalized distributions for each parameter. The retrieved median values and corresponding 1$\sigma$ uncertainties are also indicated.}
    \label{fig:hybd_corner}
\end{figure*}

\begin{figure*} [t!]\label{fig:vmr}
    \centering
    \begin{minipage}{0.49\textwidth}
        \centering
        \includegraphics[width=\textwidth]{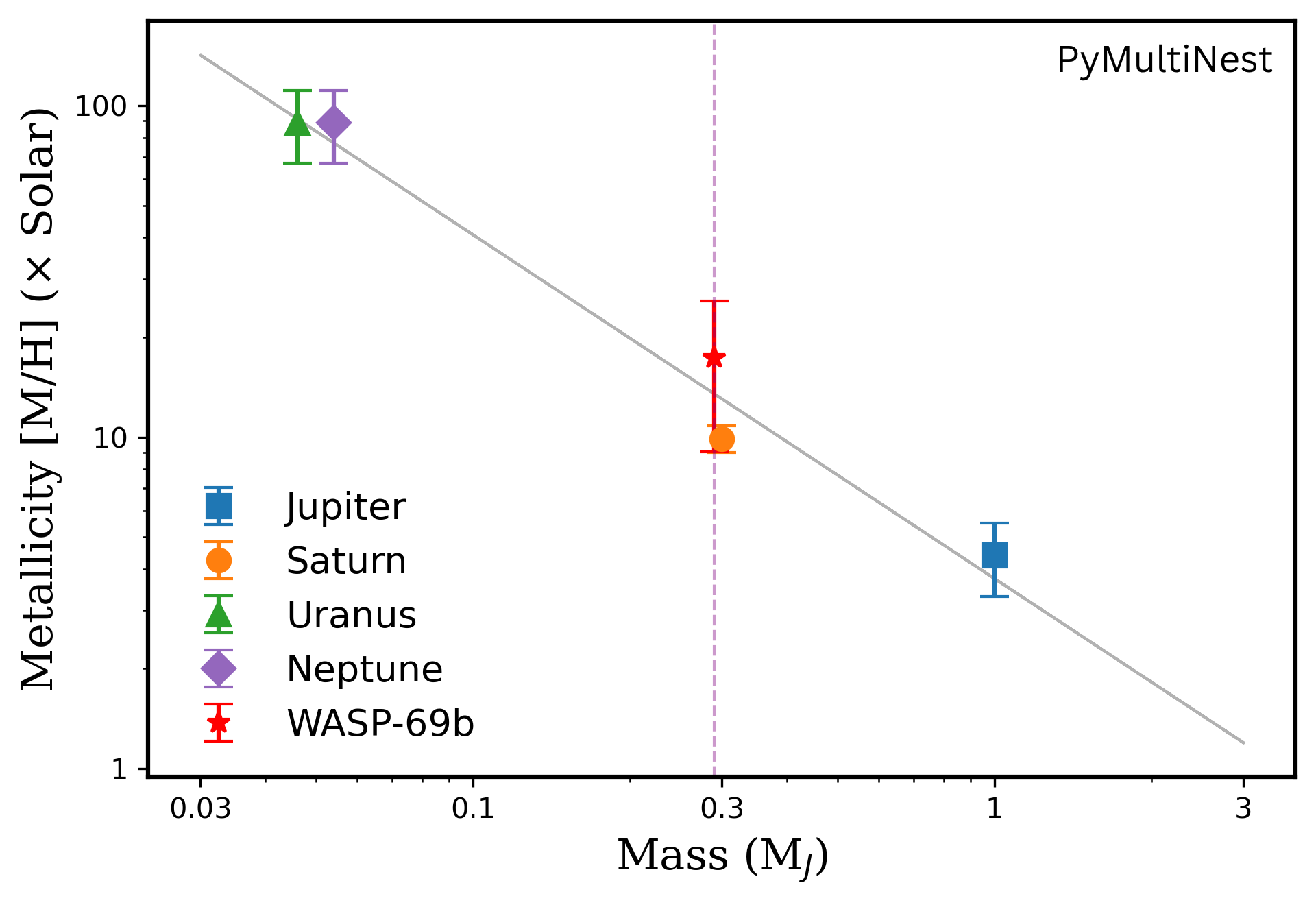}
        \parbox{0.8\linewidth}{\centering (a)}
        \label{fig:vmr_free}
        
    \end{minipage}
    \hfill
    \begin{minipage}{0.49\textwidth}
        \centering
        \includegraphics[width=\textwidth]{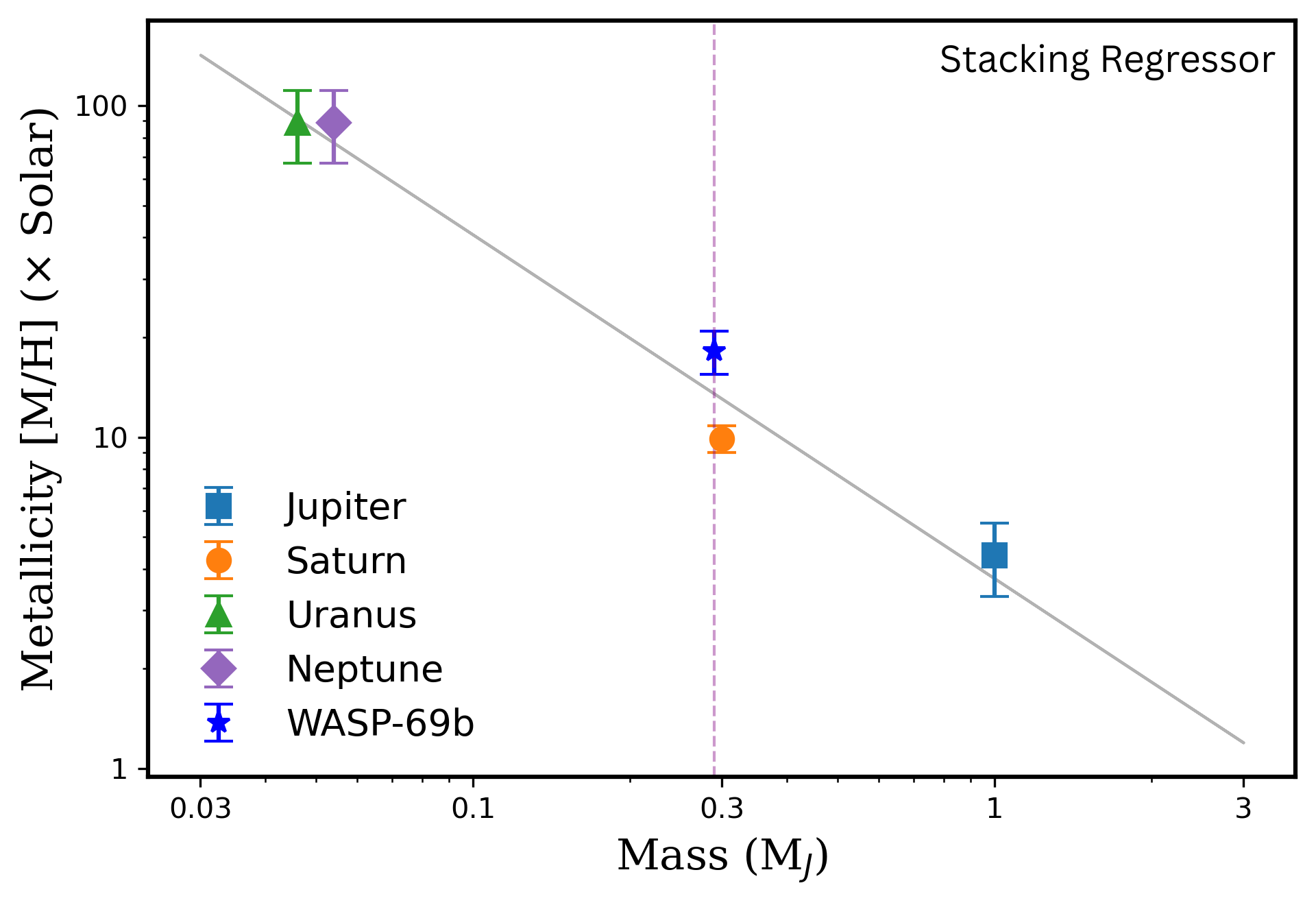}
        \parbox{0.8\linewidth}{\centering (b) }
        \label{fig:pt_free}
    \end{minipage}
       
    \caption{Retrieved atmospheric metallicity, [M/H], of WASP-69b, expressed relative to solar value, as a function of planetary mass. For comparison, the metallicities of the Solar System giant planets are shown, following \citet{Wakeford_2020}. The grey slanted line represents the expected mass-metallicity trend. The vertical dashed light purple line marks the mass of WASP-69b (0.29 M$_\mathrm{J}$). Figures (a) and (b) correspond to retrievals performed using the chemically best-fit hybrid equilibrium chemistry model, assuming non-uniform aerosol distributions in case of Bayesian and ML retrievals respectively.}
\label{fig:mass-vs-metallicity}
\end{figure*}
\subsection{\textbf{The Best-fit Model}}

\begin{figure*}[]
    \centering
    \includegraphics[width=0.85\linewidth]{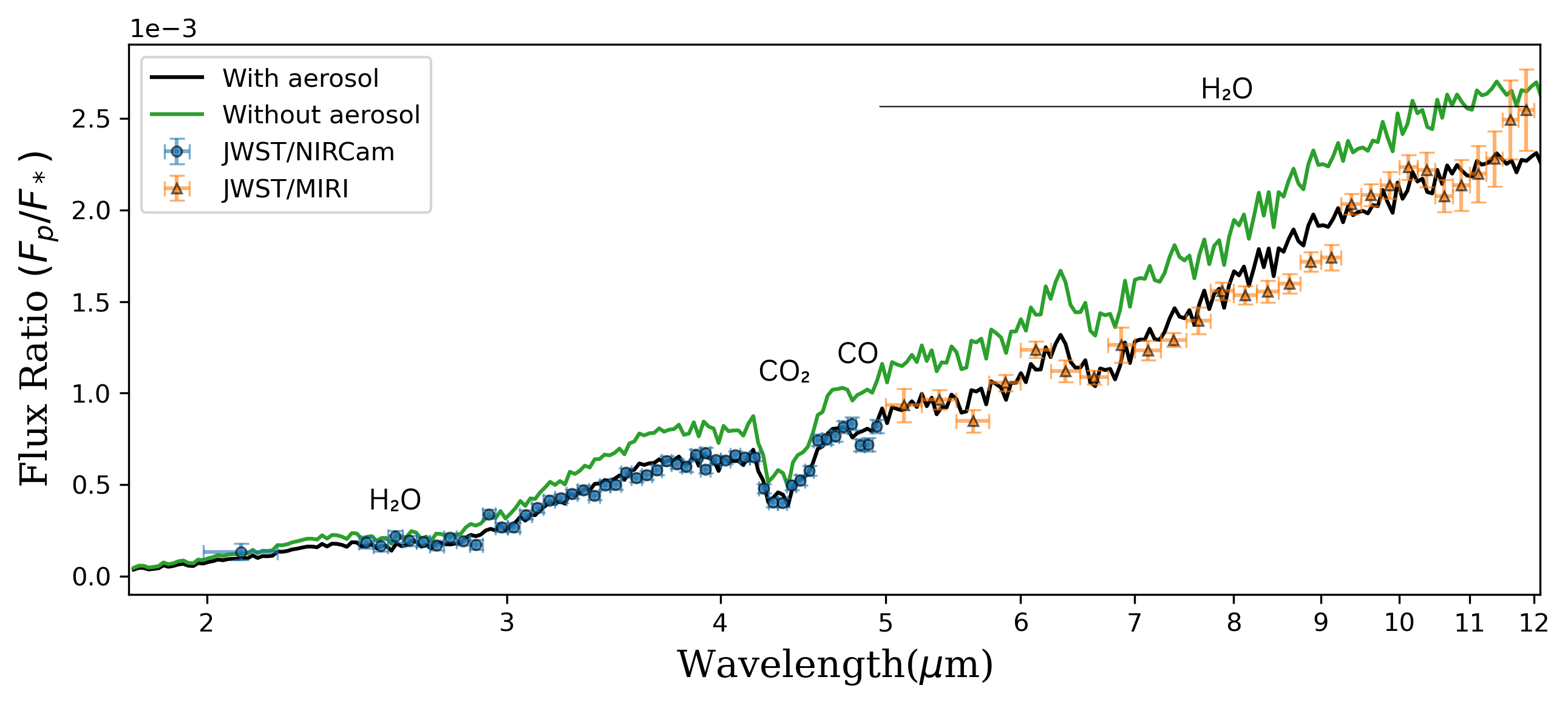}
    \label{fig:spectrum}
    \caption{Two-Stream approximation free chemistry model spectrum with and without the best-fit aerosol parameters. The spectrum shows the need for Mie scattering aerosols to fit the observations.}
    \label{aero_no_aero spectrum}
\end{figure*}

Of all the retrievals performed on the combined NIRCam and MIRI data, the free \tr{and the hybrid equilibrium chemistry} models with the presence of non-uniform aerosol contribution provides the statistically best-fit spectrum to the observations. \tr{Figure~\ref{aero_no_aero spectrum} shows the best-fit spectrum obtained from the two-stream forward model, with and without Mie-scattering MgSiO$_3$ aerosols, demonstrating the necessity of including aerosols to reproduce the observed spectrum.} 

\tr{The free chemistry model obtains} a reduced $\chi^2$ value of 2.37 corresponding to a Bayesian evidence, ln(Z) = 622.78 +/- 0.17. The best-fit spectrum for this model is shown in Figure \ref{best-fit spectrum} (a). The Bayesian corner plot for the posterior distribution of this best-fit model is also shown in Figure \ref{fig:main_corner}. \tr{On the other hand, the hybrid equilibrium chemistry} model achieves a reduced $\chi^2$ value of 2.43 and a Bayesian evidence of ln(Z) = 622.72 +/- 0.16 with a non-uniform aerosol prescription. Based on this preferred best-fit chemistry model, we can infer several key characteristics of WASP-69b's atmosphere. The hybrid chemistry model (see Section \ref{subsubsection : atmospheric profiles}), which combines equilibrium and free chemistry, retrieves a C/O ratio of $0.42^{+0.17}_{-0.13}$ and a super-solar metallicity of log[M/H] = $1.24^{+0.17}_{-0.14}$. It also suggests the presence of SO$_2$ with a VMR of log(SO$_2$) = $-4.85^{+0.28}_{-0.29}$. This supports the scenario of photochemical processing, in which SO$_2$ is produced by the oxidation of sulfur radicals released when H$_2$S is destroyed under ultraviolet irradiation \citep{tsai2023photochemically}. The model further favors the presence of high-altitude MgSiO$_3$ aerosols, with a cloud base pressure of log(P) = $-2.52^{+2.55}_{-2.12}$ and log particle sizes of $-0.23^{+0.09}_{-0.07}$ $\mu$m, covering approximately 46\% of the observable dayside disk. Such cloud properties are consistent with condensate clouds in hot Jupiter atmospheres and can effectively mute molecular absorption features \citep{mullens2024implementation}. The patchy distribution of aerosols points toward spatially varying cloud coverage, possibly induced by 3D atmospheric circulation patterns \citep{Roman_2017}. These implications, derived from the best-fit hybrid model, highlight a chemically rich and dynamically heterogeneous atmosphere for WASP-69b. The best fit spectrum for this model is shown in Figure \ref{best-fit spectrum} (b) and the the posterior distributions of the retrieved parameters are shown in Figure \ref{fig:hybd_corner} .


\tr{The elevated reduced $\chi^{2}$ values exceeding 2 obtained from all Bayesian retrieval cases using the combined NIRCam and MIRI data over the 2–12~$\mu$m range likely reflect small residual systematics and a modest underestimation of correlated uncertainties rather than a genuine model–data mismatch. Correlated (non-independent) noise components in JWST time-series spectroscopy can arise from subtle detector systematics, intra-pixel sensitivity variations, and wavelength-dependent throughput changes, even though the overall pointing stability of JWST is excellent \citep{2021AJ....161..115S}. Such correlated systematics have been directly observed in JWST exoplanet spectra \citep{holmberg2023exoplanet}, and theoretical analyses show that retrievals assuming uncorrelated Gaussian noise can overestimate the reduced $\chi^{2}$e under these conditions \citep{2021AJ....162..237I}.} 

Additionally, the machine learning retrievals using Stacking Regressor also converges on similar atmospheric properties for the hybrid equilibrium model. The machine learning retrieval recovers a C/O ratio of \tr{$0.49^{+0.01}_{-0.05}$ and a super-solar metallicity of log[M/H] = $1.26^{+0.06}_{-0.03}$. It also retrieves a log(VMR) value of $-4.99^{+0.03}_{-0.03}$} for SO$_2$, showing its possible contribution in the emission spectra.


\subsubsection{\textbf{Evaluation of Bayesian retrieval fits}}

Additionally, to quantitatively compare the performance of two best-fit retrieval models -- Model 1\textit{: free chemistry + non-uniform aerosol}, where SO$_2$ absorption is not inferred and the chemistry of H$_2$O, CO$_2$, CO, SO$_2$, and CH$_4$ is treated freely (total of 17 free parameters), and Model 2 \textit{: hybrid equilibrium + non-uniform aerosol}, where SO$_2$ (inferred) is included as a free parameter while the remaining species follow equilibrium chemistry (total of 15 free parameters) -- we employ multiple information criteria and Bayesian model selection metrics.

The first metric is the Bayesian evidence, $\mathcal{Z}$, computed using \texttt{PyMultiNest}. The logarithmic difference in evidence is $\Delta \log \mathcal{Z} = \log \mathcal{Z}_1 - \log \mathcal{Z}_2 = -0.06$. This corresponds to a Bayes factor of $\mathcal{B}_{21} \approx 1.06$, which lies well within the ``inconclusive'' regime ($|\Delta \log \mathcal{Z}| < 1$), as per the classification originally proposed by \citet{jeffreys1998theory}. Thus, there is no statistically significant preference for either model based on the Bayesian model evidence alone.

To further assess the trade-off between model complexity and goodness-of-fit, we compute the Akaike Information Criterion (AIC) \citep{akaike1974new, cavanaugh2019akaike} and Bayesian Information Criterion (BIC) \citep{schwarz1978estimating, neath2012bayesian}, defined as:

\begin{align}
\mathrm{AIC} &= 2k - 2\ln \mathcal{L}_{\max}, \label{AIC}\\
\mathrm{BIC} &= k \ln N - 2\ln \mathcal{L}_{\max}, \label{BIC}
\end{align}

where $k$ is the number of model free parameters, $N$ is the number of observed data points and $\mathcal{L}_{\max}$ is the maximum likelihood . A lower AIC or BIC implies a better model when penalizing overparameterization. 

Based on the total (NIRCam + MIRI) 79 observed data points and the obtained log‐likelihoods, the two models fit the data almost equally well. Model 1 (free chemistry, 17 parameters) has $\log\mathcal{L}_{1,\max}=0.70356$, while Model 2 (hybrid equilibrium, 15 parameters) has $\log\mathcal{L}_{2,\max}=0.69794$. The difference $\Delta\ln\mathcal{L}\approx0.0056$ (only a 0.56 increase for Model 1) is tiny given the extra two parameters in Model 1. 

Using Equations \ref{AIC} and \ref{BIC}, we get AIC$_{\text{model},1}=2(17)-2(0.70356)\approx32.59$ and AIC$_{\text{model},2}=2(15)-2(0.69794)\approx28.60$. Also, BIC$_{\text{model},1}=17\ln(79)-2(0.70356)\approx72.87$ and BIC$_{\text{model},2}=15\ln(79)-2(0.69794)\approx64.14$. The AIC and BIC values for Model 2 are both lower than those for Model 1, indicating that Model 2 provides a better balance between goodness-of-fit and model complexity. Thus, the simpler hybrid-chemistry model is statistically preferred according to both criteria.

In conclusion, while the Bayesian evidence comparison does not indicate a statistically significant preference between the free and hybrid chemistry models, the information criteria modestly favor the simpler hybrid model. Notably, SO$_2$ is retrieved only in the hybrid framework, suggesting that its spectral feature becomes prominent under chemically self-consistent assumptions for other species. This implies that the potential inference of SO$_2$ may be contingent on model assumptions and warrants cautious interpretation.

\subsubsection{\textbf{Evaluation of ML retrieval fits}}

\tr{To evaluate the machine learning retrievals, we computed the residual, the coefficient of determination ($R^2$ score), and the reduced $\chi^{2}$. Both training and testing $R^2$ scores were calculated using $k$-fold cross-validation to ensure that the models were not overfitted \citep{ghojogh2023theoryoverfittingcrossvalidation}. The corresponding values for all tested model configurations are summarized in Table~\ref{Table:all_params}. A model with lower residual and reduced $\chi^{2}$ values, and a higher $R^2$ score, provides the best spectral fit. Based on these metrics, the two-stream non-uniform aerosol model with free chemistry yields the best-fitting configuration, achieving a high $R^2$ score of 0.989 and a low residual of $0.240 \times 10^{-4}$. The model with hybrid equilibrium chemistry also performs well, ranking as the second-best fit with an $R^2$ score of 0.968 and a residual of $0.313 \times 10^{-4}$, consistent with the results from the Bayesian retrievals. Additionally, reduced $\chi^{2}$ values were computed as listed in Table~\ref{Table:all_params}. We find that the two-stream non-uniform aerosol hybrid chemistry model and the two-stream non-uniform aerosol model with free chemistry yield the lowest reduced $\chi^{2}$ values of 1.22 and 1.40, respectively, demonstrating consistency with the Bayesian retrieval results in identifying the best-fitting models. Interestingly, compared to the Bayesian goodness-of-fit, the machine-learning-based retrievals employing the stacking regressor approach yielded lower reduced $\chi^{2}$ values for the same dataset, indicating its enhanced ability to capture global spectral trends more flexibly and smoothly while minimizing small wavelength-to-wavelength residuals.}

These findings collectively indicate that both retrieval methodologies: Bayesian and machine learning, converge on metal-rich, aerosol-influenced atmospheric models as the most plausible explanation for the observed emission spectrum of WASP-69b. While the free chemistry and hybrid equilibrium models with non-uniform aerosols are identified as the best and second-best fits, respectively, additional model selection criteria such as the AIC and BIC favor the hybrid model. This preference supports the inclusion of SO$_2$, making the hybrid equilibrium scenario more physically motivated. Accordingly, in Figure~\ref{fig:mass-vs-metallicity}~(a) and (b), we present the mass–metallicity relation for WASP-69b, based on the retrieved parameters of the hybrid equilibrium model with non-uniform aerosols, as inferred from both Bayesian and machine learning retrievals. These results provide further insights into the planet’s formation and evolutionary history.

\section{\textbf{Conclusion}}\label{sec:conc}

In this study, we demonstrated the thermal emission retrieval capabilities of \texttt{NEXOTRANS} by analyzing the combined JWST NIRCam and MIRI datasets spanning 2–12 $\mu$m for WASP-69b. We conducted retrievals under both one- and two-stream radiative transfer approximations, considering scenarios of clear atmospheres and those with aerosol presence. By exploring four distinct chemistry models: free, equilibrium, hybrid, and equilibrium offset, we provided constraints on chemical abundances and especially the temperature-pressure profile, which shows great agreement among the various chemistry models (see Figure \ref{fig:vmr_pt}(b)) and possible evidence of SO$_2$ suggesting the presence of photochemical processes in the dayside atmosphere of WASP-69b. The key findings and conclusions of this analysis are summarized below:

\begin{enumerate}

    \item The abundances of key molecular species--H$_2$O, CO$_2$, CO, CH$_4$, and SO$_2$--are well constrained by the retrievals. Among these, H$_2$O and CO$_2$ are found to be the dominant contributors to the observed absorption features. The results also indicate no significant detection of CH$_4$ in the thermal emission spectrum, suggesting a low abundance or absence of methane on the dayside of WASP-69b. Possible contribution due to SO$_2$ in the 7-8 $\mu$m region is also seen among the explored models.
    
    \item The retrieved carbon-to-oxygen (C/O) ratios span a wide range, from sub-solar to super-solar values, depending on the cloud treatment adopted in the models. For scenarios with uniform aerosol coverage, the median C/O ratios range between 0.30 and 0.73, while models incorporating non-uniform MgSiO$_3$ aerosols yield slightly higher values, ranging from 0.42 to 0.83. In contrast, clear atmosphere models produce C/O values between 0.29 and 0.57. These findings are broadly consistent with those of \citet{schlawin2024multiple}, who report a wider possible ranges for the C/O ratio, extending from approximately 0.26 to 0.94 depending on the model.

    \item The retrieved metallicity [M/H], also vary depending on the assumed atmospheric scenario. For models incorporating MgSiO$_3$ aerosols, the metallicity is found to lie between $\sim$ 6.6 and 19.05 times the solar value in the Bayesian retrievals and 7.7--39.8 times in the ML retrievals. In the case of a clear atmosphere, [M/H] spans from $\sim$ 6.3 to 19.5 times solar including both Bayesian and ML retrievals. These estimates are generally consistent with those reported by \citet{schlawin2024multiple}, who find a metallicity range of approximately 4 to 14 times solar when considering all model scenarios, also including their less favored scattering model.
   
    \item  Among all the explored models, the free chemistry model with non-uniform aerosol coverage provides the best statistical fit (reduced $\chi^2 \approx 2.37$) to the combined JWST NIRCam and MIRI datasets. The best-fit spectrum (Figure \ref{best-fit spectrum} (a)) shows that the observed absorption features are best explained by H$_2$O and CO$_2$ with log(vmr) of $-3.78^{+0.15}_{-0.17}$ and $-5.77^{+0.09}_{-0.10}$ respectively. Although not clearly distinguishable, the CO abundances are also inferred at a moderate log(vmr) of $-3.78^{+0.21}_{-0.22}$.
    
    \item Among the models utilizing the \texttt{NEXOCHEM} chemistry grid, the one that best-fits the global data is the hybrid equilibrium chemistry model with presence of non-uniform aerosol coverage. This is also the second best statistical model with a reduced $\chi^2$ value of 2.43. Information criteria calculation using metrics such as the AIC and BIC favor this model, indicating it achieves a more optimal trade-off between goodness-of-fit and model complexity.
    
    \item The retrieved VMR profiles (Figure \ref{fig:vmr_pt} (a)) indicate that the abundances of H$_2$O, CO$_2$, CO, and CH$_4$ are only slightly shifted from the equilibrium abundances when assuming NEXOTRANS's approximate disequilibrium chemical modeling approaches. The higher SO$_2$ abundance in the hybrid and equilibrium offset models suggests that if present, photochemical processes are active on the dayside of WASP-69b.

    \item The free chemistry retrievals on just the MIRI dataset from Section \ref{validation_section} suggests that the abundances of H$_2$O are significantly overestimated in that case, as compared to the combined NIRCam and MIRI retrievals. The MIRI only retrievals obtain log(vmr) between -1.64 and -1.83 whereas the combined retrievals has values ranging between -3.78 and -4.69. This underscores the necessity of spectroscopic observations spanning wide wavelength ranges to robustly constrain the abundances of key atmospheric constituents.

    \item \tr{The retrieved T–P profiles for WASP-69~b, shown in Figure~\ref{fig:vmr_pt}(b), display a smooth decrease in temperature with altitude, ranging from approximately 1400~K in the deeper atmosphere to about 500~K in the upper layers, and reveal an almost isothermal profile in the deeper atmospheric layers. \citet{schlawin2024multiple} reported similar T–P profiles for their scattering and cloud-layer models, with temperatures near 600~K at the top and increasing to $\geq$1400~K at the bottom, along with a deep isothermal region beginning around P~$>$~10$^{-3}$~bar. Importantly, these T–P profiles remain consistent across the different equilibrium-based chemical modeling approaches explored in this study. This consistency indicates that the inferred thermal structure is a robust feature of WASP-69~b’s atmosphere, largely insensitive to the specific chemical or aerosol assumptions adopted in the retrievals.}

    \item The retrieved T–P profile for the best-fit free chemistry model shows an overall lower atmospheric temperature \tr{compared to the other chemistry models. In the free chemistry case, the assumption of vertically constant VMRs decouples the molecular abundances from the temperature structure. Without equilibrium constraints to guide the vertical variation of species, the retrieval compensates by adjusting the temperature, leading to a cooler thermal profile relative to the equilibrium-based models.}

    \item  The retrieved aerosol properties show that non-uniform cloud treatments favor larger MgSiO$_3$ particles (up to $\sim 5\,\mu$m) and higher-altitude cloud decks (with median base pressures around log(P) $\sim -1.80$ to $-2.63$~bar), suggesting enhanced vertical mixing or coagulation in spatially heterogeneous cloud regions. Overall, in all the models explored, non-uniform aerosols are favored when considered (see the reduced $\chi^2$ values in Table \ref{tab:cloud_params_ML_corrected}).

    \item Best-fit analysis of all the retrievals performed using both Bayesian and machine learning methods indicate the combined observations of WASP-69b from both NIRCam and MIRI are best explained by an atmosphere of super-solar metallicity and the presence of clouds or more specifically MgSiO$_3$ aerosol condensates, along with possible contribution of photochemical species such as SO$_2$. These results demonstrate the unique capability of NEXOTRANS to constrain the atmospheric composition, thermal structure and infer possible influence of disequilibrium chemistry on the observed spectrum by utilizing diverse sets of models and retrieval techniques.

\end{enumerate}

\begin{acknowledgments}
L.M. acknowledges financial support from DAE, Government of India, for this work. L.M. also gratefully acknowledges support from Breakthrough Listen at the University of Oxford through a sub-award at NISER under Agreement R82799/CN013, provided as part of a global collaboration under the Breakthrough Listen project funded by the Breakthrough Prize Foundation.

\vspace{5mm}
\facilities{JWST}

\software{NEXOTRANS \citep{deka2025}, POSEIDON \citep{2023JOSS....8.4873M}, Python \citep{10.5555/1593511}, numba \citep{lam2015numba}, matplotlib \citep{Hunter:2007}, mpi4py \citep{dalcin2005mpi, dalcin2008mpi, dalcin2019mpi, dalcin2021mpi4py, rogowski2022mpi4py}, Sci-kit learn \citep{scikit-learn}.}

\end{acknowledgments}

\section{Appendix} \label{appendix}
\subsection{\tr{\textbf{Uncertainty calculation in ML}}}

\tr{In this section, we discuss the calculation of the $\sigma$ confidence levels. In the machine learning model described in Section~\ref{subsection:retrieval framework}, we employ a \texttt{stacking regressor} that combines \texttt{random forest}, \texttt{gradient boosting}, and \texttt{k-nearest neighbor} as base models, with a \texttt{ridge regressor} serving as the meta-model. This algorithm provides only point predictions, which makes it challenging to derive posterior distributions and, consequently, to estimate parameter uncertainties. To determine the $\pm1\sigma$ interval (i.e., the 68$\%$ confidence region), a distribution of parameter values is required.}

\tr{To construct this parameter distribution, we perturb the observed transit depths by introducing a 10\% random error and iteratively sample model predictions within the $\pm10\%$ uncertainty range of the data. This procedure generates a distribution of parameter values corresponding to the perturbed observational space, thereby approximating the confidence region around the retrieved solution.}

\tr{Let the free parameters be denoted as $\theta = (\theta_{1}, \theta_{2}, \theta_{3}, \ldots, \theta_{n})$. There are $N$ samples for each parameter, forming a parameter space defined as 
$$ S = \{\theta^{(1)}, \theta^{(2)}, \ldots, \theta^{(N)}\}. $$
Here, $S$ represents the joint probability density function $P(\theta|D)$, where $D$ is the observed spectrum.}

\tr{The best-fit value for each parameter is taken as the median (50th percentile), and the 1$\sigma_{i}$ confidence interval is defined between the 16th and 84th percentiles of the distribution of $\theta_{i}$. Thus, the upper and lower bounds of the 1$\sigma_{i}$ errors are calculated as 
$$ \sigma_{i}^{+} = q_{0.84} - q_{0.50}, \quad \sigma_{i}^{-} = q_{0.50} - q_{0.16}, $$
where $q_{n}$ represents the $n$-th quantile.}

\subsection{\tr{\textbf{Calculation of reduced $\chi^{2}$ score in ML}}}
\tr{The reduced chi-square for the ML model is calculated as}
\begin{equation}
\tr{\chi_{\nu}^{2} = \frac{1}{\text{dof}} 
    \sum_{i=1}^{N} 
    \left( 
        \frac{T^{\text{model}}_i - T^{\text{observed}}_i}{T^{\text{err}}_i} 
    \right)^{2}},
\end{equation}

\tr{where \textit{dof} (degrees of freedom) is the number of transit depth points in the observation minus the number of parameters retrieved by the model. $T_i^{\text{model}}$ is the model-predicted transit depth, binned and interpolated to match the wavelength grid of the observed spectrum. $T_i^{\text{observed}}$ is the observed JWST transit depth, and $T_i^{\text{err}}$ is its corresponding uncertainty. The resulting $\chi^{2}$ values are reported in Table~\ref{Table:all_params}.}

\tr{The computation of $\chi^{2}$ here differs slightly from that used in the Bayesian retrieval, where it is calculated as}
\[
\tr{\chi^2_{\text{best}} = -2\left(\ln \mathcal{L}_{\text{max}} - \text{normalized-likelihood}\right)},
\]
\tr{with $\mathcal{L}_{\text{max}}$ representing the maximum likelihood obtained from the nested sampling algorithm, and}
\[
\tr{\text{normalized-likelihood} = -\frac{1}{2} \sum_{i=1}^{N} 
\ln\!\left( 2\pi\, \sigma_{\mathrm{eff},i}^2 \right)},
\]
\tr{where $\sigma_{\mathrm{eff},i}$ denotes the effective uncertainty in the transit depth.}

\clearpage
\bibliography{main}{}

\begin{thebibliography}{}
\expandafter\ifx\csname natexlab\endcsname\relax\def\natexlab#1{#1}\fi
\providecommand{\url}[1]{\href{#1}{#1}}
\providecommand{\dodoi}[1]{doi:~\href{http://doi.org/#1}{\nolinkurl{#1}}}
\providecommand{\doeprint}[1]{\href{http://ascl.net/#1}{\nolinkurl{http://ascl.net/#1}}}
\providecommand{\doarXiv}[1]{\href{https://arxiv.org/abs/#1}{\nolinkurl{https://arxiv.org/abs/#1}}}

\bibitem[{Ackerman \& Marley(2001)}]{ackerman2001precipitating}
Ackerman, A.~S., \& Marley, M.~S. 2001, The Astrophysical Journal, 556, 872

\bibitem[{Akaike(1974)}]{akaike1974new}
Akaike, H. 1974, IEEE transactions on automatic control, 19, 716

\bibitem[{August {et~al.}(2023)August, Bean, Zhang, Lunine, Xue, Line, \&
  Smith}]{august2023confirmation}
August, P.~C., Bean, J.~L., Zhang, M., {et~al.} 2023, The Astrophysical Journal
  Letters, 953, L24

\bibitem[{Birkby(2018)}]{birkby2018exoplanet}
Birkby, J.~L. 2018, arXiv preprint arXiv:1806.04617

\bibitem[{Breiman(2001)}]{breiman2001random}
Breiman, L. 2001, Machine learning, 45, 5

\bibitem[{Buchner(2021)}]{buchner2021ultranest}
Buchner, J. 2021, arXiv preprint arXiv:2101.09604

\bibitem[{Buchner {et~al.}(2014)Buchner, Georgakakis, Nandra, Hsu, Rangel,
  Brightman, Merloni, Salvato, Donley, \& Kocevski}]{buchner2014x}
Buchner, J., Georgakakis, A., Nandra, K., {et~al.} 2014, Astronomy \&
  Astrophysics, 564, A125

\bibitem[{Cartier {et~al.}(2016)Cartier, Beatty, Zhao, Line, Ngo, Mawet,
  Stassun, Wright, Kreidberg, Fortney, {et~al.}}]{cartier2016near}
Cartier, K.~M., Beatty, T.~G., Zhao, M., {et~al.} 2016, The Astronomical
  Journal, 153, 34

\bibitem[{Castelli \& Kurucz(2004)}]{castelli2004new}
Castelli, F., \& Kurucz, R.~L. 2004, arXiv preprint astro-ph/0405087

\bibitem[{Cavanaugh \& Neath(2019)}]{cavanaugh2019akaike}
Cavanaugh, J.~E., \& Neath, A.~A. 2019, Wiley Interdisciplinary Reviews:
  Computational Statistics, 11, e1460

\bibitem[{Changeat \& Edwards(2021)}]{changeat2021hubble}
Changeat, Q., \& Edwards, B. 2021, The Astrophysical Journal Letters, 907, L22

\bibitem[{Charbonneau {et~al.}(2008)Charbonneau, Knutson, Barman, Allen, Mayor,
  Megeath, Queloz, \& Udry}]{charbonneau2008broadband}
Charbonneau, D., Knutson, H.~A., Barman, T., {et~al.} 2008, The Astrophysical
  Journal, 686, 1341

\bibitem[{Charnay {et~al.}(2015)Charnay, Meadows, \& Leconte}]{Charnay_2015}
Charnay, B., Meadows, V., \& Leconte, J. 2015, The Astrophysical Journal, 813,
  15, \dodoi{10.1088/0004-637x/813/1/15}

\bibitem[{Constantinou \& Madhusudhan(2024)}]{Constantinou}
Constantinou, S., \& Madhusudhan, N. 2024, MNRAS, 530, 3252–3277

\bibitem[{Cover \& Hart(1967)}]{cover1967nearest}
Cover, T., \& Hart, P. 1967, IEEE transactions on information theory, 13, 21

\bibitem[{Cubillos \& Blecic(2021)}]{cubillos2021pyrat}
Cubillos, P.~E., \& Blecic, J. 2021, Monthly Notices of the Royal Astronomical
  Society, 505, 2675

\bibitem[{Dalcin(2019)}]{dalcin2019mpi}
Dalcin, L. 2019, MPI for Python,  Feb

\bibitem[{Dalcin \& Fang(2021)}]{dalcin2021mpi4py}
Dalcin, L., \& Fang, Y.-L.~L. 2021, Computing in Science \& Engineering, 23, 47

\bibitem[{Dalc{\'\i}n {et~al.}(2005)Dalc{\'\i}n, Paz, \&
  Storti}]{dalcin2005mpi}
Dalc{\'\i}n, L., Paz, R., \& Storti, M. 2005, Journal of Parallel and
  Distributed Computing, 65, 1108

\bibitem[{Dalc{\'\i}n {et~al.}(2008)Dalc{\'\i}n, Paz, Storti, \&
  D’El{\'\i}a}]{dalcin2008mpi}
Dalc{\'\i}n, L., Paz, R., Storti, M., \& D’El{\'\i}a, J. 2008, Journal of
  Parallel and Distributed Computing, 68, 655

\bibitem[{De~Kok {et~al.}(2011)De~Kok, Helling, Stam, Woitke, \&
  Witte}]{de2011influence}
De~Kok, R.~J., Helling, C., Stam, D.~M., Woitke, P., \& Witte, S. 2011,
  Astronomy \& Astrophysics, 531, A67

\bibitem[{{Deka} {et~al.}(2025){Deka}, {Khan}, {Dewan}, {Ghosh}, {Das}, \&
  {Majumdar}}]{deka2025}
{Deka}, T., {Khan}, T.~B., {Dewan}, S., {et~al.} 2025, \apj, 989, 50,
  \dodoi{10.3847/1538-4357/add33d}

\bibitem[{Deming {et~al.}(2006)Deming, Harrington, Seager, \&
  Richardson}]{deming2006strong}
Deming, D., Harrington, J., Seager, S., \& Richardson, L.~J. 2006, The
  Astrophysical Journal, 644, 560

\bibitem[{Deming \& Knutson(2020)}]{deming2020highlights}
Deming, D., \& Knutson, H.~A. 2020, Nature Astronomy, 4, 453

\bibitem[{Deming {et~al.}(2005)Deming, Seager, Richardson, \&
  Harrington}]{Deming2005}
Deming, D., Seager, S., Richardson, L.~J., \& Harrington, J. 2005, Nature, 434,
  740, \dodoi{10.1038/nature03507}

\bibitem[{Foote {et~al.}(2021)Foote, Lewis, Kilpatrick, Goyal, Bruno, Wakeford,
  Robbins-Blanch, Kataria, MacDonald, L{\'o}pez-Morales,
  {et~al.}}]{foote2021emission}
Foote, T.~O., Lewis, N.~K., Kilpatrick, B.~M., {et~al.} 2021, The Astronomical
  Journal, 163, 7

\bibitem[{Fortney(2018)}]{fortney2018modeling}
Fortney, J.~J. 2018, Astrophysics of Exoplanetary Atmospheres: 2nd Advanced
  School on Exoplanetary Science, 51

\bibitem[{France {et~al.}(2010)France, Stocke, Yang, Linsky, Wolven, Froning,
  Green, \& Osterman}]{France_2010}
France, K., Stocke, J.~T., Yang, H., {et~al.} 2010, The Astrophysical Journal,
  712, 1277–1286, \dodoi{10.1088/0004-637x/712/2/1277}

\bibitem[{Friedman(2001)}]{friedman2001greedy}
Friedman, J.~H. 2001, Annals of statistics, 1189

\bibitem[{Gandhi \& Madhusudhan(2018)}]{gandhi2018retrieval}
Gandhi, S., \& Madhusudhan, N. 2018, Monthly Notices of the Royal Astronomical
  Society, 474, 271

\bibitem[{Gandhi \& Madhusudhan(2019)}]{gandhi2019new}
---. 2019, arXiv preprint arXiv:1903.11603

\bibitem[{Gebhard {et~al.}(2024)Gebhard, Wildberger, Dax, Kofler, Angerhausen,
  Quanz, \& Sch{\"o}lkopf}]{Gebhard2024}
Gebhard, T.~D., Wildberger, J., Dax, M., {et~al.} 2024, arXiv e-prints.
\newblock \doarXiv{2410.21477}

\bibitem[{Gebhard {et~al.}(2025)Gebhard, Wildberger, Dax, Kofler, Angerhausen,
  Quanz, \& Sch{\"o}lkopf}]{gebhard2025flow}
---. 2025, Astronomy \& Astrophysics, 693, A42

\bibitem[{Ghojogh \&
  Crowley(2023)}]{ghojogh2023theoryoverfittingcrossvalidation}
Ghojogh, B., \& Crowley, M. 2023, The Theory Behind Overfitting, Cross
  Validation, Regularization, Bagging, and Boosting: Tutorial.
\newblock \doarXiv{1905.12787}

\bibitem[{Glidic {et~al.}(2022)Glidic, Schlawin, Wiser, Zhou, Deming, \&
  Line}]{glidic2022atmospheric}
Glidic, K., Schlawin, E., Wiser, L., {et~al.} 2022, The Astronomical Journal,
  164, 19

\bibitem[{Greene {et~al.}(2023)Greene, Bell, Ducrot, Dyrek, Lagage, \&
  Fortney}]{greene2023thermal}
Greene, T.~P., Bell, T.~J., Ducrot, E., {et~al.} 2023, Nature, 618, 39

\bibitem[{Grillmair {et~al.}(2007)Grillmair, Charbonneau, Burrows, Armus,
  Stauffer, Meadows, Van~Cleve, \& Levine}]{grillmair2007spitzer}
Grillmair, C.~J., Charbonneau, D., Burrows, A., {et~al.} 2007, The
  Astrophysical Journal, 658, L115

\bibitem[{Guillot(2010)}]{guillot2010radiative}
Guillot, T. 2010, Astronomy \& Astrophysics, 520, A27

\bibitem[{Holmberg \& Madhusudhan(2023)}]{holmberg2023exoplanet}
Holmberg, M., \& Madhusudhan, N. 2023, Monthly Notices of the Royal
  Astronomical Society, 524, 377

\bibitem[{Hoogkamer {et~al.}(2025)Hoogkamer, Kini, Salmi, Watts, \&
  Buchner}]{cp8c-2nbk}
Hoogkamer, M., Kini, Y., Salmi, T., Watts, A.~L., \& Buchner, J. 2025, Phys.
  Rev. D, 112, 023008, \dodoi{10.1103/cp8c-2nbk}

\bibitem[{Hunter(2007)}]{Hunter:2007}
Hunter, J.~D. 2007, Computing in Science \& Engineering, 9, 90,
  \dodoi{10.1109/MCSE.2007.55}

\bibitem[{Husser {et~al.}(2013)Husser, Wende-von Berg, Dreizler, Homeier,
  Reiners, Barman, \& Hauschildt}]{husser2013new}
Husser, T.-O., Wende-von Berg, S., Dreizler, S., {et~al.} 2013, Astronomy \&
  Astrophysics, 553, A6

\bibitem[{{Ih} \& {Kempton}(2021)}]{2021AJ....162..237I}
{Ih}, J., \& {Kempton}, E. M.~R. 2021, \aj, 162, 237,
  \dodoi{10.3847/1538-3881/ac173b}

\bibitem[{Jeffreys(1998)}]{jeffreys1998theory}
Jeffreys, H. 1998, The theory of probability (OuP Oxford)

\bibitem[{Kawahara {et~al.}(2022)Kawahara, Kawashima, Masuda, Crossfield,
  van~den Bekerom, Kitzmann, Morris, Pannier, Nugroho, \&
  Ishikawa}]{kawahara2022exojax}
Kawahara, H., Kawashima, Y., Masuda, K., {et~al.} 2022, Astrophysics Source
  Code Library, ascl

\bibitem[{Kitzmann {et~al.}(2020)Kitzmann, Heng, Oreshenko, Grimm, Apai,
  Bowler, Burgasser, \& Marley}]{kitzmann2020helios}
Kitzmann, D., Heng, K., Oreshenko, M., {et~al.} 2020, The Astrophysical
  Journal, 890, 174

\bibitem[{Kurucz \& Peytremann(1975)}]{kurucz1975table}
Kurucz, R.~L., \& Peytremann, E. 1975, SAO Special Report\# 362, part 1., 362

\bibitem[{Lam {et~al.}(2015)Lam, Pitrou, \& Seibert}]{lam2015numba}
Lam, S.~K., Pitrou, A., \& Seibert, S. 2015, in Proceedings of the Second
  Workshop on the LLVM Compiler Infrastructure in HPC, 1--6

\bibitem[{Lavie {et~al.}(2017)Lavie, Mendon{\c{c}}a, Mordasini, Malik,
  Bonnefoy, Demory, Oreshenko, Grimm, Ehrenreich, \& Heng}]{lavie2017helios}
Lavie, B., Mendon{\c{c}}a, J.~M., Mordasini, C., {et~al.} 2017, The
  Astronomical Journal, 154, 91

\bibitem[{Lee {et~al.}(2012)Lee, Fletcher, \& Irwin}]{lee2012optimal}
Lee, J.-M., Fletcher, L.~N., \& Irwin, P.~G. 2012, Monthly Notices of the Royal
  Astronomical Society, 420, 170

\bibitem[{Line \& Parmentier(2016)}]{line2016influence}
Line, M.~R., \& Parmentier, V. 2016, The Astrophysical Journal, 820, 78

\bibitem[{Line {et~al.}(2013)Line, Wolf, Zhang, Knutson, Kammer, Ellison,
  Deroo, Crisp, \& Yung}]{line2013systematic}
Line, M.~R., Wolf, A.~S., Zhang, X., {et~al.} 2013, The Astrophysical Journal,
  775, 137

\bibitem[{{MacDonald}(2023)}]{2023JOSS....8.4873M}
{MacDonald}, R.~J. 2023, The Journal of Open Source Software, 8, 4873,
  \dodoi{10.21105/joss.04873}

\bibitem[{MacDonald \& Batalha(2023{\natexlab{a}})}]{macdonald2023catalog}
MacDonald, R.~J., \& Batalha, N.~E. 2023{\natexlab{a}}, Research Notes of the
  AAS, 7, 54

\bibitem[{MacDonald \& Batalha(2023{\natexlab{b}})}]{MacDonald_2023}
---. 2023{\natexlab{b}}, Research Notes of the AAS, 7, 54,
  \dodoi{10.3847/2515-5172/acc46a}

\bibitem[{Madhusudhan(2019)}]{Madhusudhan_2019}
Madhusudhan, N. 2019, Annual Review of Astronomy and Astrophysics, 57, 617,
  \dodoi{https://doi.org/10.1146/annurev-astro-081817-051846}

\bibitem[{Madhusudhan \& Seager(2009)}]{madhusudhan2009temperature}
Madhusudhan, N., \& Seager, S. 2009, The Astrophysical Journal, 707, 24

\bibitem[{Madhusudhan \& Seager(2010)}]{madhusudhan2010inference}
---. 2010, The Astrophysical Journal, 725, 261

\bibitem[{Mansfield(2023)}]{mansfield2023revealing}
Mansfield, M. 2023, Astrophysics and Space Science, 368, 24

\bibitem[{Mikal-Evans {et~al.}(2019)Mikal-Evans, Sing, Goyal, Drummond, Carter,
  Henry, Wakeford, Lewis, Marley, Tremblin, {et~al.}}]{mikal2019emission}
Mikal-Evans, T., Sing, D.~K., Goyal, J.~M., {et~al.} 2019, Monthly Notices of
  the Royal Astronomical Society, 488, 2222

\bibitem[{Min {et~al.}(2020)Min, Ormel, Chubb, Helling, \&
  Kawashima}]{min2020arcis}
Min, M., Ormel, C.~W., Chubb, K., Helling, C., \& Kawashima, Y. 2020, Astronomy
  \& Astrophysics, 642, A28

\bibitem[{Molli{\`e}re {et~al.}(2019)Molli{\`e}re, Wardenier, Van~Boekel,
  Henning, Molaverdikhani, \& Snellen}]{molliere2019petitradtrans}
Molli{\`e}re, P., Wardenier, J., Van~Boekel, R., {et~al.} 2019, Astronomy \&
  Astrophysics, 627, A67

\bibitem[{Mukherjee {et~al.}(2023)Mukherjee, Batalha, Fortney, \&
  Marley}]{mukherjee2023picaso}
Mukherjee, S., Batalha, N.~E., Fortney, J.~J., \& Marley, M.~S. 2023, The
  Astrophysical Journal, 942, 71

\bibitem[{Mukherjee {et~al.}(2025)Mukherjee, Schlawin, Bell, Fortney, Beatty,
  Greene, Ohno, Murphy, Parmentier, Line, {et~al.}}]{mukherjee2025jwst}
Mukherjee, S., Schlawin, E., Bell, T.~J., {et~al.} 2025, The Astrophysical
  Journal Letters, 982, L39

\bibitem[{Mullens {et~al.}(2024)Mullens, Lewis, \&
  MacDonald}]{mullens2024implementation}
Mullens, E., Lewis, N.~K., \& MacDonald, R.~J. 2024, The Astrophysical Journal,
  977, 105

\bibitem[{Neath \& Cavanaugh(2012)}]{neath2012bayesian}
Neath, A.~A., \& Cavanaugh, J.~E. 2012, Wiley Interdisciplinary Reviews:
  Computational Statistics, 4, 199

\bibitem[{Pedregosa {et~al.}(2011)Pedregosa, Varoquaux, Gramfort, Michel,
  Thirion, Grisel, Blondel, Prettenhofer, Weiss, Dubourg, Vanderplas, Passos,
  Cournapeau, Brucher, Perrot, \& Duchesnay}]{scikit-learn}
Pedregosa, F., Varoquaux, G., Gramfort, A., {et~al.} 2011, Journal of Machine
  Learning Research, 12, 2825

\bibitem[{Pinhas \& Madhusudhan(2017)}]{pinhas2017signatures}
Pinhas, A., \& Madhusudhan, N. 2017, Monthly Notices of the Royal Astronomical
  Society, 471, 4355

\bibitem[{Robinson \& Salvador(2023)}]{robinson2023exploring}
Robinson, T.~D., \& Salvador, A. 2023, The Planetary Science Journal, 4, 10

\bibitem[{Rogowski {et~al.}(2022)Rogowski, Aseeri, Keyes, \&
  Dalcin}]{rogowski2022mpi4py}
Rogowski, M., Aseeri, S., Keyes, D., \& Dalcin, L. 2022, IEEE Transactions on
  Parallel and Distributed Systems, 34, 611

\bibitem[{Roman \& Rauscher(2017)}]{Roman_2017}
Roman, M., \& Rauscher, E. 2017, The Astrophysical Journal, 850, 17,
  \dodoi{10.3847/1538-4357/aa8ee4}

\bibitem[{{Schlawin} {et~al.}(2021){Schlawin}, {Leisenring}, {McElwain},
  {Misselt}, {Don}, {Greene}, {Beatty}, {Nikolov}, {Kelly}, \&
  {Rieke}}]{2021AJ....161..115S}
{Schlawin}, E., {Leisenring}, J., {McElwain}, M.~W., {et~al.} 2021, \aj, 161,
  115, \dodoi{10.3847/1538-3881/abd8d4}

\bibitem[{Schlawin {et~al.}(2024)Schlawin, Mukherjee, Ohno, Bell, Beatty,
  Greene, Line, Challener, Parmentier, Fortney,
  {et~al.}}]{schlawin2024multiple}
Schlawin, E., Mukherjee, S., Ohno, K., {et~al.} 2024, The Astronomical Journal,
  168, 104

\bibitem[{Schwarz(1978)}]{schwarz1978estimating}
Schwarz, G. 1978, The annals of statistics, 461

\bibitem[{Spiegel \& Burrows(2013)}]{spiegel2013thermal}
Spiegel, D.~S., \& Burrows, A. 2013, The Astrophysical Journal, 772, 76

\bibitem[{Stevenson {et~al.}(2014)Stevenson, Bean, Madhusudhan, \&
  Harrington}]{stevenson2014deciphering}
Stevenson, K.~B., Bean, J.~L., Madhusudhan, N., \& Harrington, J. 2014, The
  Astrophysical Journal, 791, 36

\bibitem[{Toon {et~al.}(1989)Toon, McKay, Ackerman, \&
  Santhanam}]{toon1989rapid}
Toon, O.~B., McKay, C., Ackerman, T., \& Santhanam, K. 1989, Journal of
  Geophysical Research: Atmospheres, 94, 16287

\bibitem[{Tsai {et~al.}(2023)Tsai, Lee, Powell, Gao, Zhang, Moses, H{\'e}brard,
  Venot, Parmentier, Jordan, {et~al.}}]{tsai2023photochemically}
Tsai, S.-M., Lee, E.~K., Powell, D., {et~al.} 2023, Nature, 617, 483

\bibitem[{Van~Rossum \& Drake(2009)}]{10.5555/1593511}
Van~Rossum, G., \& Drake, F.~L. 2009, Python 3 Reference Manual (Scotts Valley,
  CA: CreateSpace)

\bibitem[{Vasist {et~al.}(2023)Vasist, Rozet, Absil, Mollière, Nasedkin, \&
  Louppe}]{Vasist2023}
Vasist, M., Rozet, F., Absil, O., {et~al.} 2023, Astronomy \& Astrophysics,
  672, A147, \dodoi{10.1051/0004-6361/202245263}

\bibitem[{Wachiraphan {et~al.}(2024)Wachiraphan, Berta-Thompson, Diamond-Lowe,
  Winters, Murray, Zhang, Xue, Morley, Rosario-Franco, \&
  Duvvuri}]{wachiraphan2024thermal}
Wachiraphan, P., Berta-Thompson, Z.~K., Diamond-Lowe, H., {et~al.} 2024, arXiv
  preprint arXiv:2410.10987

\bibitem[{Wakeford \& Dalba(2020)}]{Wakeford_2020}
Wakeford, H.~R., \& Dalba, P.~A. 2020, Philosophical Transactions of the Royal
  Society A: Mathematical, Physical and Engineering Sciences, 378, 20200054,
  \dodoi{10.1098/rsta.2020.0054}

\bibitem[{Waldmann {et~al.}(2015)Waldmann, Tinetti, Rocchetto, Barton,
  Yurchenko, \& Tennyson}]{waldmann2015tau}
Waldmann, I.~P., Tinetti, G., Rocchetto, M., {et~al.} 2015, The Astrophysical
  Journal, 802, 107

\bibitem[{Xue {et~al.}(2024)Xue, Bean, Zhang, Mahajan, Ih, Eastman, Lunine,
  Mansfield, Coy, Kempton, {et~al.}}]{xue2024jwstemission}
Xue, Q., Bean, J.~L., Zhang, M., {et~al.} 2024, The Astrophysical Journal
  Letters, 973, L8

\bibitem[{Yip {et~al.}(2022)Yip, Changeat, Al-Refaie, \& Waldmann}]{Yip2022}
Yip, K.~H., Changeat, Q., Al-Refaie, A., \& Waldmann, I.~P. 2022, arXiv
  e-prints.
\newblock \doarXiv{2205.07037}

\bibitem[{Zahnle {et~al.}(2009)Zahnle, Marley, Freedman, Lodders, \&
  Fortney}]{Zahnle_2009}
Zahnle, K., Marley, M.~S., Freedman, R.~S., Lodders, K., \& Fortney, J.~J.
  2009, The Astrophysical Journal, 701, L20–L24,
  \dodoi{10.1088/0004-637x/701/1/l20}

\end{thebibliography}
\bibliographystyle{aasjournal}

\end{document}